\documentclass[12pt]{article}

\newenvironment{wileykeywords}{\textsf{Keywords:}\hspace{\stretch{1}}}{\hspace{\stretch{1}}\rule{1ex}{1ex}}

\usepackage{amsmath,amssymb}
\usepackage{graphicx}% Include figure files
\usepackage{color}% Include colors for document elements
\usepackage{dcolumn}% Align table columns on decimal point
\usepackage{bm}% bold math
\usepackage[numbers,super,comma,sort&compress]{natbib}

\definecolor{background-color}{gray}{0.98}
\definecolor{background-color}{gray}{0.98}

\title{The Behavior of Cyclohexane Confined in Slit Carbon Nanopore}
\author{Yu. D. Fomin(1,2), E.N. Tsiok(1) and V.N. Ryzhov(1,2) \thanks{(1)Institute for High Pressure Physics Russian Academy of Sciences,
(2) Moscow Institute of Physics and Technology (State
University)}}
\begin{document}

\maketitle

\begin{abstract}
It is well known that confining a liquid into a pore strongly
alters the liquid behavior. Investigations of the effect of
confinement are of great importance for many scientific and
technological applications. Here we present a molecular dynamics
study of the behavior of cyclohexane confined in carbon slit
pores. The local structure and orientational ordering of
cyclohexane molecules are investigated. It is shown that the
system freezes with decreasing the pore width, and the freezing
temperature of nanoconfined cyclohenae is higher than the bulk
one.
\end{abstract}

\begin{wileykeywords}
Cyclohexane, slit pore, graphite walls, local structure,
orientational ordering, freezing.
%A list of five key words or phrases which best characterize the paper are required for indexing.
\end{wileykeywords}

\clearpage

%*****************Graphical Table of Contents******************** THIS IS MANDATORY *******************

%\begin{figure}[h]
%\centering
%\colorbox{background-color}{
%\fbox{
%\begin{minipage}{1.0\textwidth}
%\includegraphics[width=55mm,height=50mm]{H106-abstract.eps} % Pick only one of the two styles by uncommenting the corresponding \includegraphics
%\includegraphics[width=110mm,height=20mm]{cc.eps}
%\\
%Benzene in graphite and amorphous carbon slit pores is studied. It
%is shown that in graphite pore the benzene molecules arrange
%parallel to the surface in small enough pores while in amorphous
%carbon pores the benzene molecules do not observe any strong
%ordering even in small pores.
%\end{minipage}
%}}
%\end{figure}

% makes references listed with 1., 2., etc.
  \makeatletter
  \renewcommand\@biblabel[1]{#1.}
  \makeatother

\bibliographystyle{apsrev}

\renewcommand{\baselinestretch}{1.5}
\normalsize

\clearpage

%\section*{\sffamily \Large SUMMARY}

%A short summary of the main contributions in the paper is required. This summary should be carefully prepared for it is automatically the source for most abstracts.

%It is well known that confining a liquid into a pore strongly
%alters the liquid behavior. Investigations of the effect of
%confinement are of great importance for many scientific and
%technological applications. Here we present a study of the
%behavior of benzene confined in carbon slit pores. Two types of
%pores are considered - graphite and amorphous carbon ones. We show
%that the effect of different pore structure is of crucial
%importance for the benzene behavior: while in graphite pores
%benzene molecules arrange parallel to the walls there is no clear
%ordering in amorphous carbon pores.

\section*{\sffamily \Large I INTRODUCTION} % Not needed for rapid communications

It is well known that the nature of spatial ordering of a
molecular system depends on the dimensionality of the space to
which it is confined. Confinement can modify the dynamics, the
thermodynamics and the structural properties of liquids when the
range of molecular interaction and the length scale associated
with position correlations in a system are similar to the length
scale of confinement \cite{rice}. As a result, fluids confined in
a pore demonstrate the behavior strongly different from the bulk
case. In slit pores the melting temperature  depends strongly on
the pore width $H$ and the nature of the fluid-wall and
fluid-fluid attractive interaction \cite{rad}. The parameter $H$
determines not only the magnitude of the shift of the melting
temperature but also the structure of confined fluids and
appearance of the possible new phases. The density profile becomes
modulated unlike the isotropic one in bulk liquids.
%The diffusion
%coefficient usually drops under confinement, etc.
The study of different nanoconfined systems becomes important for
both fundamental research and technology because there are a lot
of real physical and biological phenomena and processes that
depend on the properties of such systems. Nanoconfinement is
considerably interesting also due to the new physics observed in
these systems.
%These phenomena becomes important for both
%fundamental research and technology in such fields as cohesion,
%nanotribology and so on.

In spite of  great importance of this issue, there is a lack of
systematic investigations of effect of confinement on the behavior
of aromatic hydrocarbons. Only the case of benzene received a lot
of attention in the literature
\cite{be1,be2,be3,pibond,pibond1,pibond2,tworings,shim,benzen-me,benzene-nanotube}.
However, liquids with more complex chemical structure are also of
interest.

In this paper, we present a systematic molecular dynamics study of
cyclohexane in the slit carbon pores. In our previous publications
we considered the behavior of benzene in slit pores with graphite
and amorphous carbon walls \cite{benzen-me} and inside an armchair
nanotube \cite{benzene-nanotube}. Benzene has chemical formula
$C_6H_{6}$ and planar molecule. The chemical formula of
cyclohexane is $C_6H_{12}$. Adding six more hydrogens changes the
bonding inside the carbon ring and therefore alters the
interaction of the ring with external molecules. Moreover, the
molecule changes its shape: it is not a planar, but warped. In
this respect it is interesting to compare the behavior of benzene
and cyclohexane in the same kinds of pores. Three types of
differences can affect the results: the different bonding inside
the molecules, the different geometric shape of the molecules and
the different number of hydrogens.

The planar structure of the benzene molecule corresponds to the
in-plane $sp^2$ hybrid orbitals of $C$, which produce a resonant
bond shared by all $C$ atoms. The cyclohexane molecule has six
$sp^3$ carbon atoms. The strong $\pi$-electron interaction of
benzene with nanopore surfaces was found
\cite{pibond,pibond1,pibond2,doss} which plays an essential role
in the formation of the local structure of nanoconfined benzene
\cite{benzen-me,benzene-nanotube}. At the same time the stronger
cyclohexane absorption on graphite than benzene absorption was
observed \cite{absor}, and one can suppose that the interaction of
$sp^3$ carbon atoms of cyclohexane with the graphite confining
walls can lead to different structure and thermodynamic behavior
of cyclohexane in comparison with benzene.

The behavior of cyclohexane in nanopores is interesting in its own
rights. In particular, in Ref. \cite{c6h12-nmr} it was proposed to
use cyclohexane for calibration of NMR experiments which can be
important for development of experimental techniques. Several
publications described the behavior of cyclohexane in mica pores
\cite{mica1,mica3}. For example, in Refs.
\cite{mica01,mica2,mica02} the crystallization of cyclohexane
confined between parallel mica surfaces was studied. It was found
that with decreasing the pore width up to about $4 nm$ (6 or 7
layers of cyclohexane) the crystallization occurs which is
observed from the sudden increase of a yield stress. It is
interesting that it was found the remarkable increase of the
melting temperature in comparison with the bulk case. In
experiments \cite{mica01,mica2,mica02} the solidification was
observed at pressures close to the atmospheric ones, which gives
the possibility to suppose that the geometric confinement alone
can lead to the solidlike behavior. However, in Ref.
\cite{christenson} this result was opposed, because in this work
the experiment was repeated, and no evidence of the increase of
melting temperature was found. The qualitative support for the
results of Refs. \cite{mica01,mica2,mica02} was obtained from
computer simulations of model systems \cite{cui,ravi}.

An analysis of Ref. \cite{ravi} suggests that the sign of the
change of the melting temperature under confinement depends on the
ratio of the fluid-wall to the fluid-fluid attractive interaction.
A number of studies of different systems confined within activated
carbon fibers show that the freezing temperature can increase or
decrease depending on the nature of the system. In Ref.
\cite{benz} it was found an increase about $65 K$ in the freezing
temperature of benzene in activated carbons. On the other hand,
for water, a considerable decrease of the freezing temperature was
observed \cite{water,water1}.

It is interesting to note that in some cases the intermediate
hexatic phase can exist \cite{ravi,water}. As it was proposed in
Ref. \cite{ravi} the hexatic phase can be observed for the small
number of layers. For example, in $CCl_4$ the hexatic phase was
found for two layers \cite{ravi}. In the case of water the hexatic
phase does exist in the case of strictly two-dimensional model
\cite{dfrt1,dfrt2,dfrt3} which demonstrates waterlike behavior
\cite{jcp2008,wepre,we_inv}.

In the present paper, we study the local structure and freezing
transition of system of cyclohexane molecules confined in graphite
slit pores using the realistic AIREBO potential \cite{airebo}. The
molecular models and simulation methods are described in Sec. II,
and the results are presented in Sec. III and discussed in Sec.
IV.

\section*{\sffamily \Large II METHODOLOGY}

%((Place Computational Methods here. Not needed for review articles))

%((Computational results should be prepared following the IUPAC guidelines (See Journal of Computational Chemistry, 20: 1587-1590 and 20:1591-1592). In particular, it is required that the level of theory employed is appropriate to the problem at hand, and that the sufficient details about methodology are provided to allow the work to be reproduced.)

%((In full papers, this section appears immediately after the introduction. In Rapid Communications, this section appears just before the Acknowledgments.))

\begin{figure}
\includegraphics[width=7cm, height=7cm]{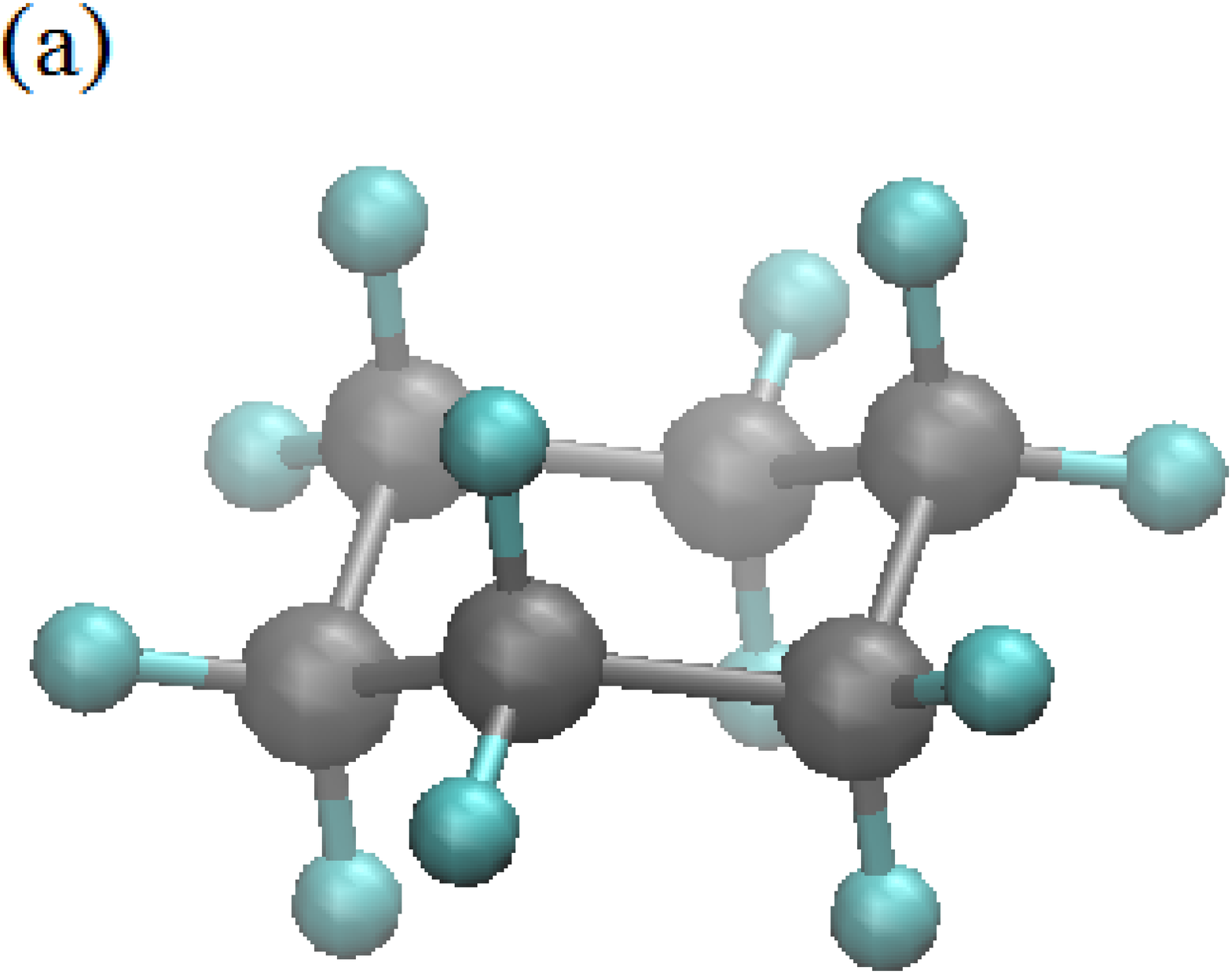}%
\includegraphics[width=7cm, height=7cm]{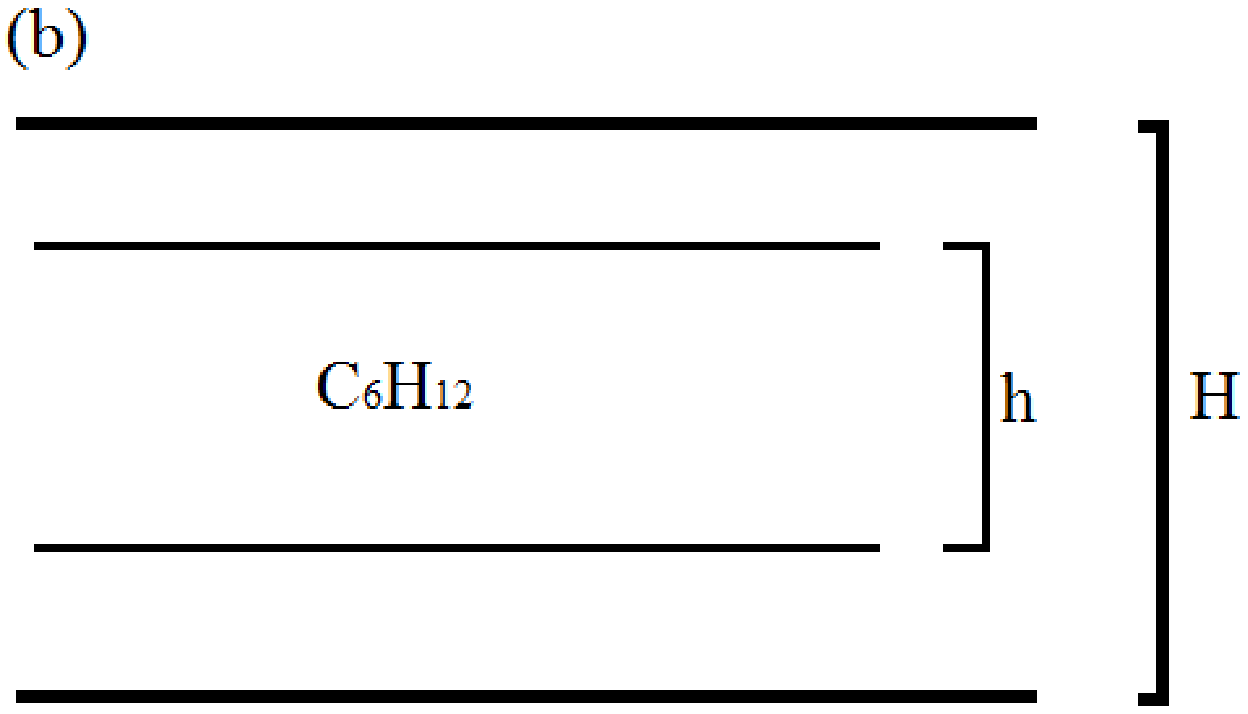}%

\caption{\label{fig:fig0} (a) Cyclohexane molecule. Larger balls
represent carbon atoms while the smaller ones - hydrogens. (b) A
cartoon of the simulation box (see the text).}
\end{figure}

We start with brief description of cyclohexane molecule. It
consists of a six carbon ring with two hydrogen atoms bonded to
each of the ring sites (Fig.~\ref{fig:fig0}). Unlike benzene, the
carbon ring of cyclohexane is not flat, but warped. The total size
of the molecule is approximately $5.8 \AA$.

The simulation methodology is similar to the one used in our
previous publication for benzene \cite{benzen-me}. We studied
$200$ molecules of cyclohexane in slit pores. The walls of the
pores consisted of two graphite layers. The inner layer was mobile
while the outer one was held rigid in order to stabilize the
system. The distance between the outer graphite layers is denoted
as $H$. The inner layers are separated from the outer ones by $3.4
\AA$. We denote the distance between the inner layers as $h$.
Obviously, $h=(H - 6.8) \AA$. The parameter $h$ determines the
area in $z$ direction available for cyclohexane molecules.

The system was simulated by molecular dynamics method in $NVT$
ensemble (constant number of particles, volume and temperature).
The temperature was maintained at $T=300K$ by means of Nose-Hoover
thermostat. The timestep was $dt_{gr}=0.001$ fs. The system was
equilibrated for $1.5 \cdot 10^6$ steps. Then $5.0 \cdot 10^6$
steps were done in order to calculate the averages.

In order to see the effect of the walls separation on the
structure of the liquid we simulate the system at different pore
heights. The system sizes in $x$ and $y$ direction are
$L_x=49.1902 \AA$ and $L_y=46.86 \AA$. The size in $z$ direction
(from the bottom of the lower wall up to the top of the upper
wall) changes from $H^{min}=19.355 \AA$ up to $H^{max}=33.355 \AA$
with step $dH=2.0 \AA$. In total, $8$ different pore sizes were
studied.

The system was periodic in $XY$ plane, but confined along $z$ axis
which is perpendicular to the walls.

All interactions in the system were modelled by AIREBO potential
\cite{airebo}. This potential was specially developed for
simulation of hydrocarbon systems. It includes carbon-carbon,
carbon-hydrogen and hydrogen-hydrogen interactions, i.e. all
interactions involved in the systems under investigation.

All simulations reported in this article were done in lammps
simulation package \cite{lammps}.

\section*{\sffamily \Large III RESULTS}

%((Place Results here. Not needed for review articles.))

%((Equations should be inserted using standard LaTeX equation and eqnarray environments, not as graphics, and should be set in the main text))
%Equation                                           (1)
%((References should be superscripted and appear after punctuation.1,2 Please define all acronyms at their first usage except IR, UV, NMR, and DNA or similar commonly understood terms.))

\begin{figure}
\includegraphics[width=5cm, height=5cm]{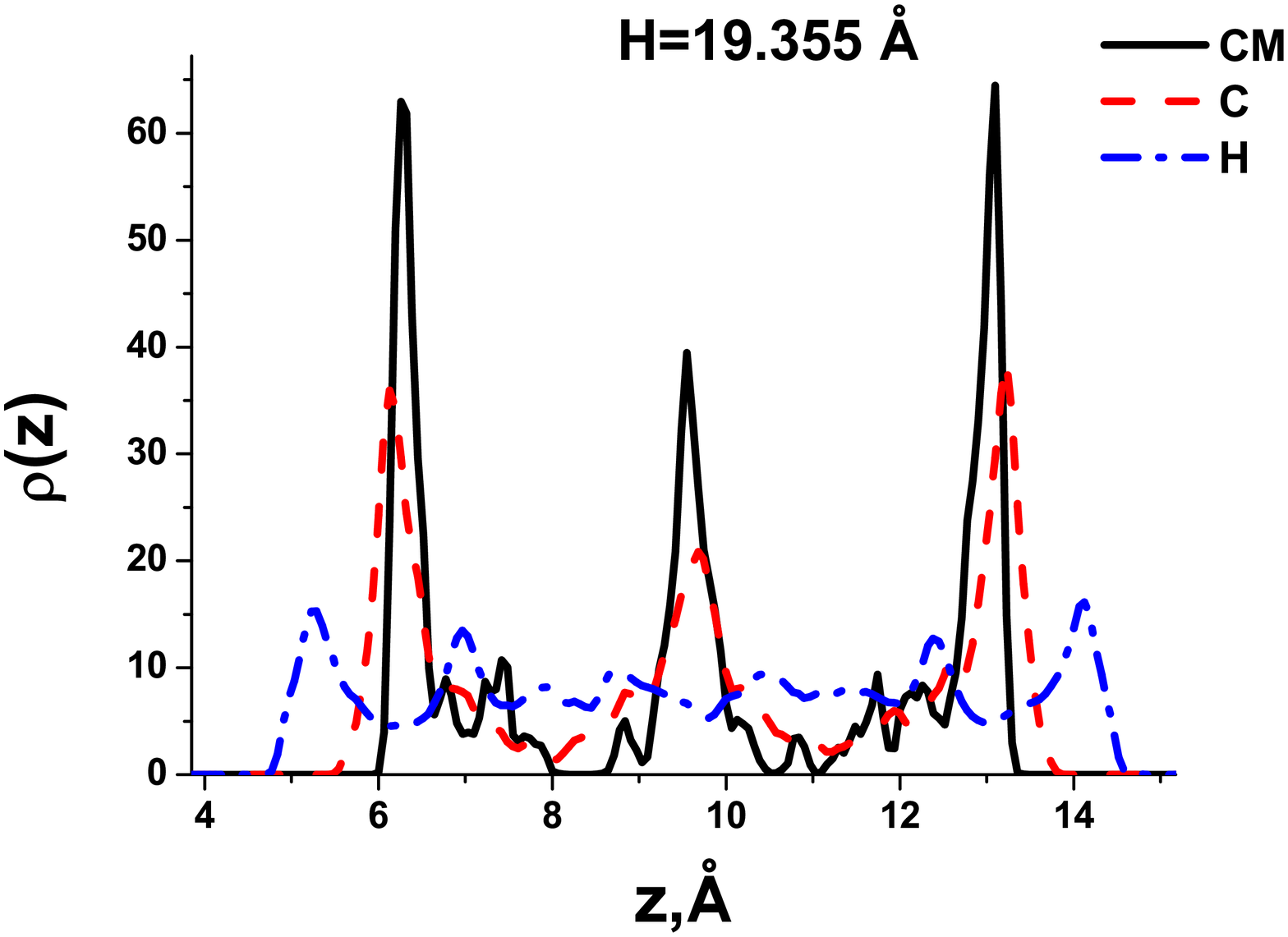}%
\includegraphics[width=5cm, height=5cm]{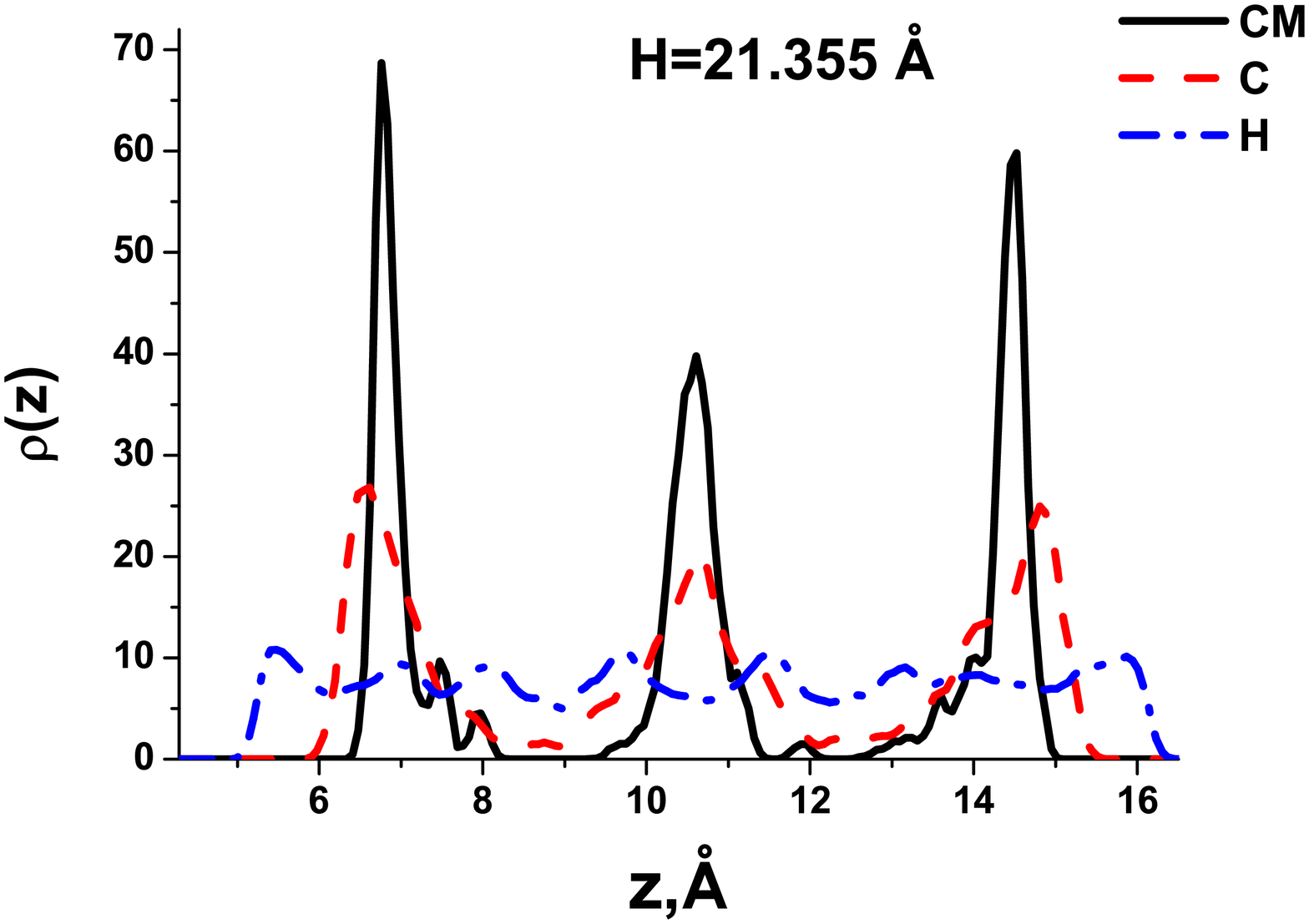}%

\includegraphics[width=5cm, height=5cm]{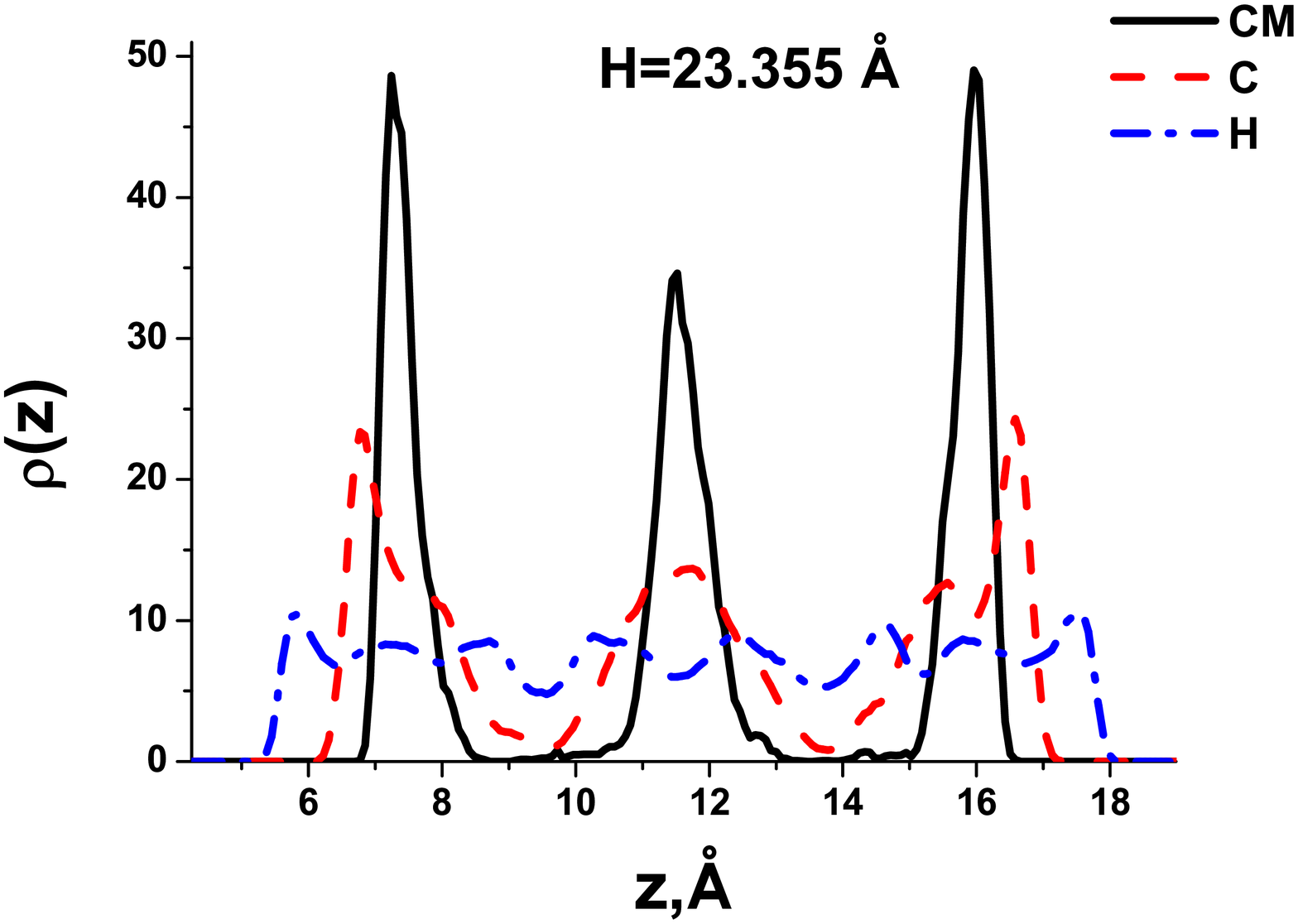}%
\includegraphics[width=5cm, height=5cm]{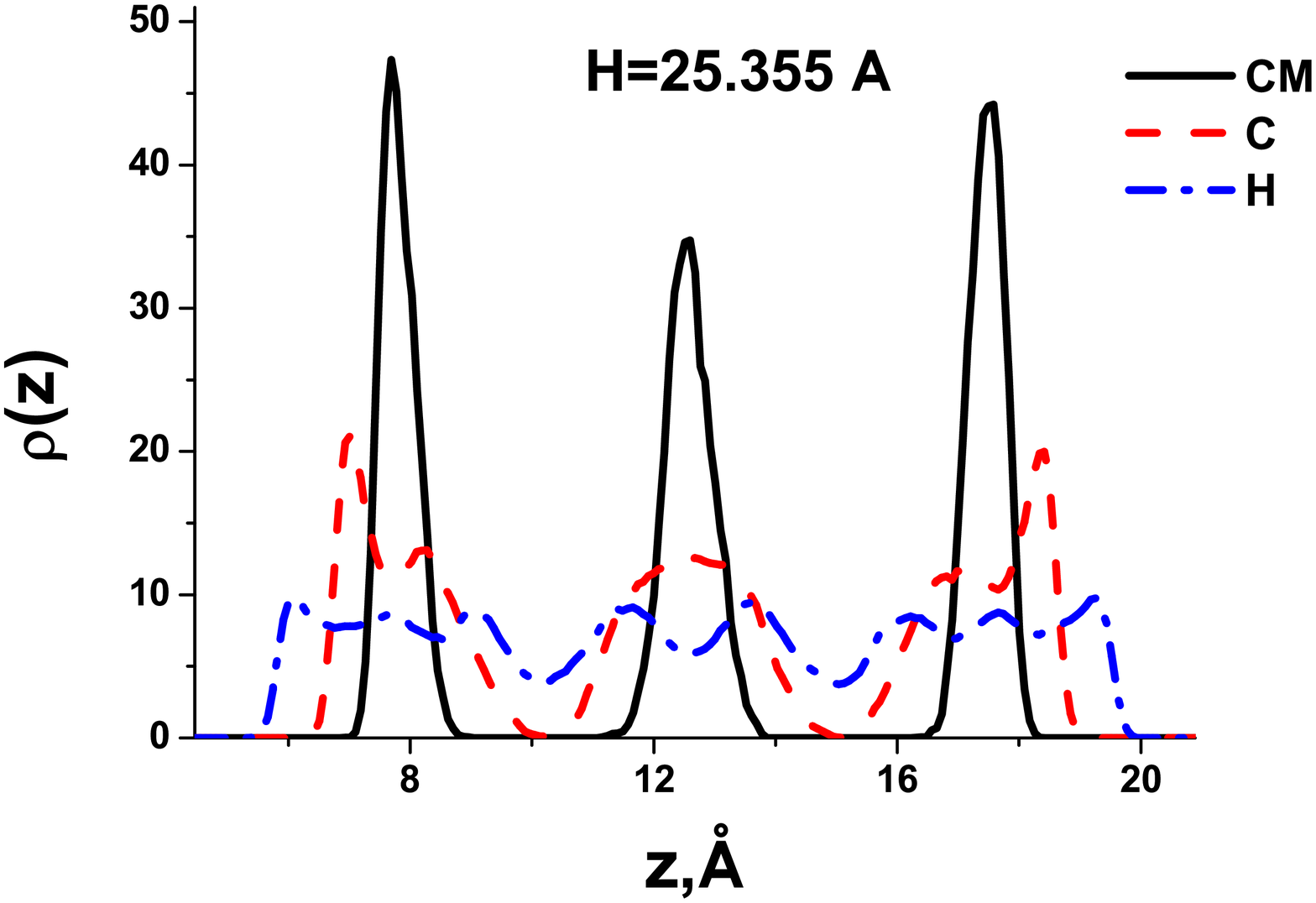}%

\includegraphics[width=5cm, height=5cm]{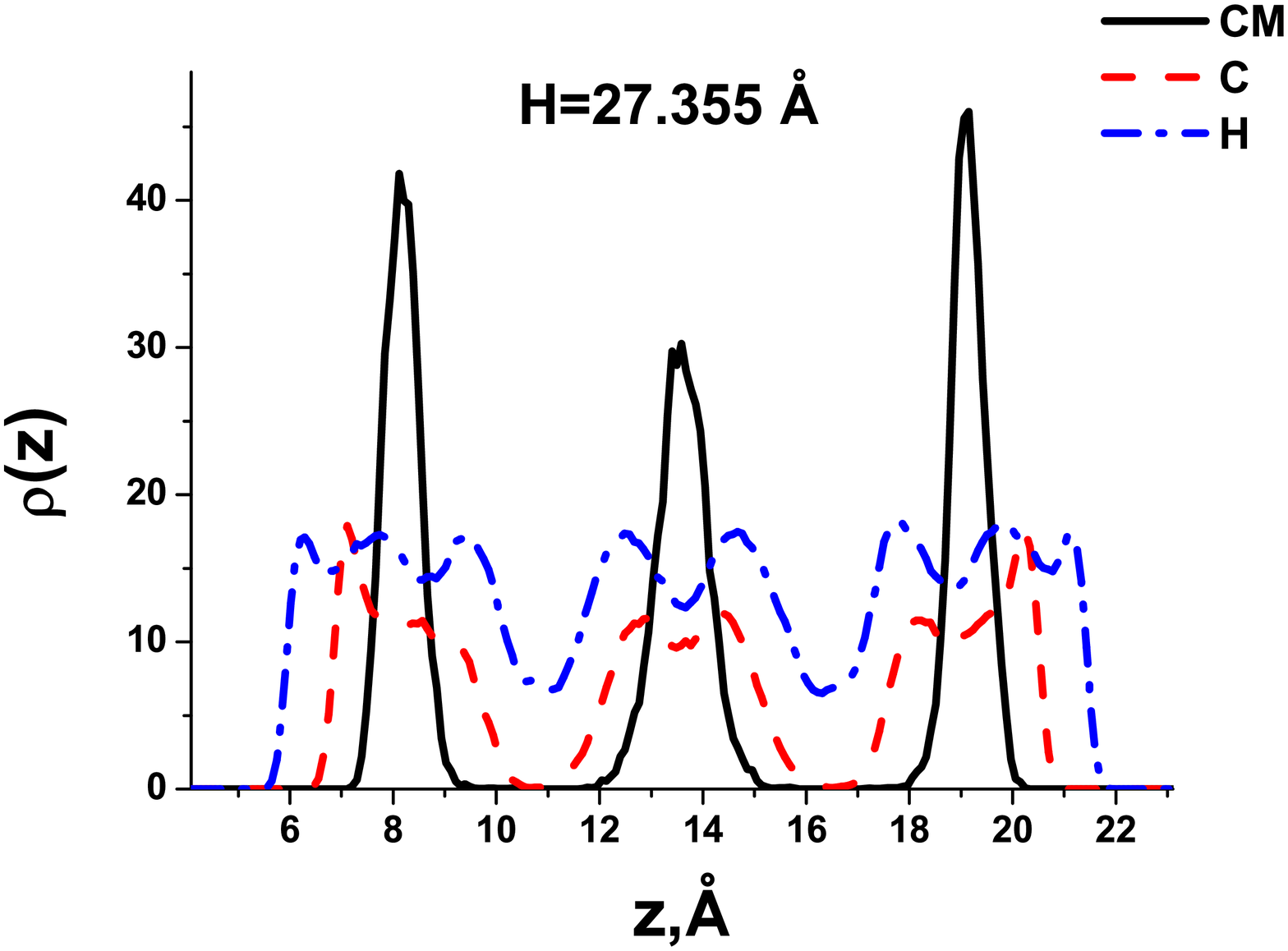}%
\includegraphics[width=5cm, height=5cm]{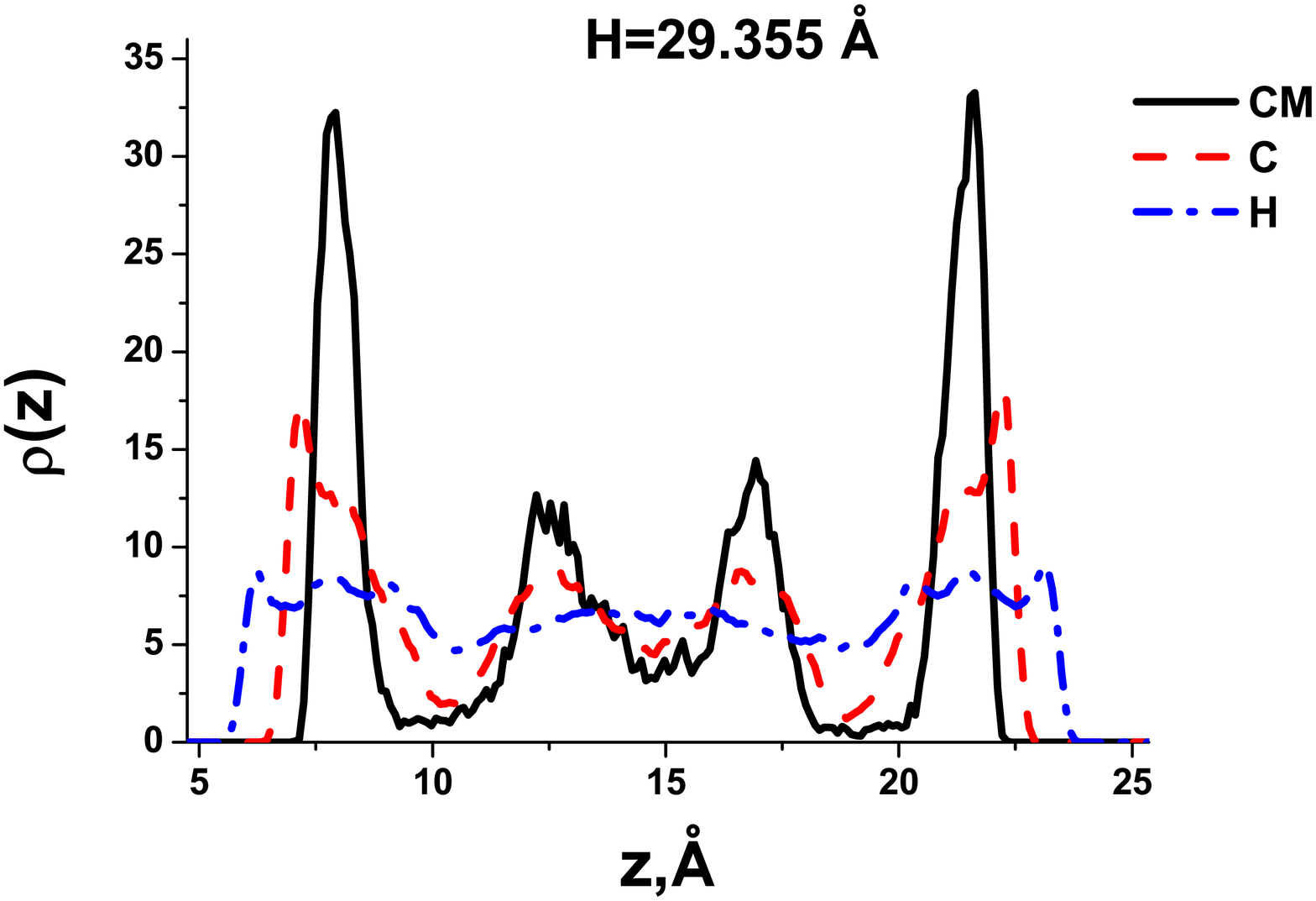}%

\includegraphics[width=5cm, height=5cm]{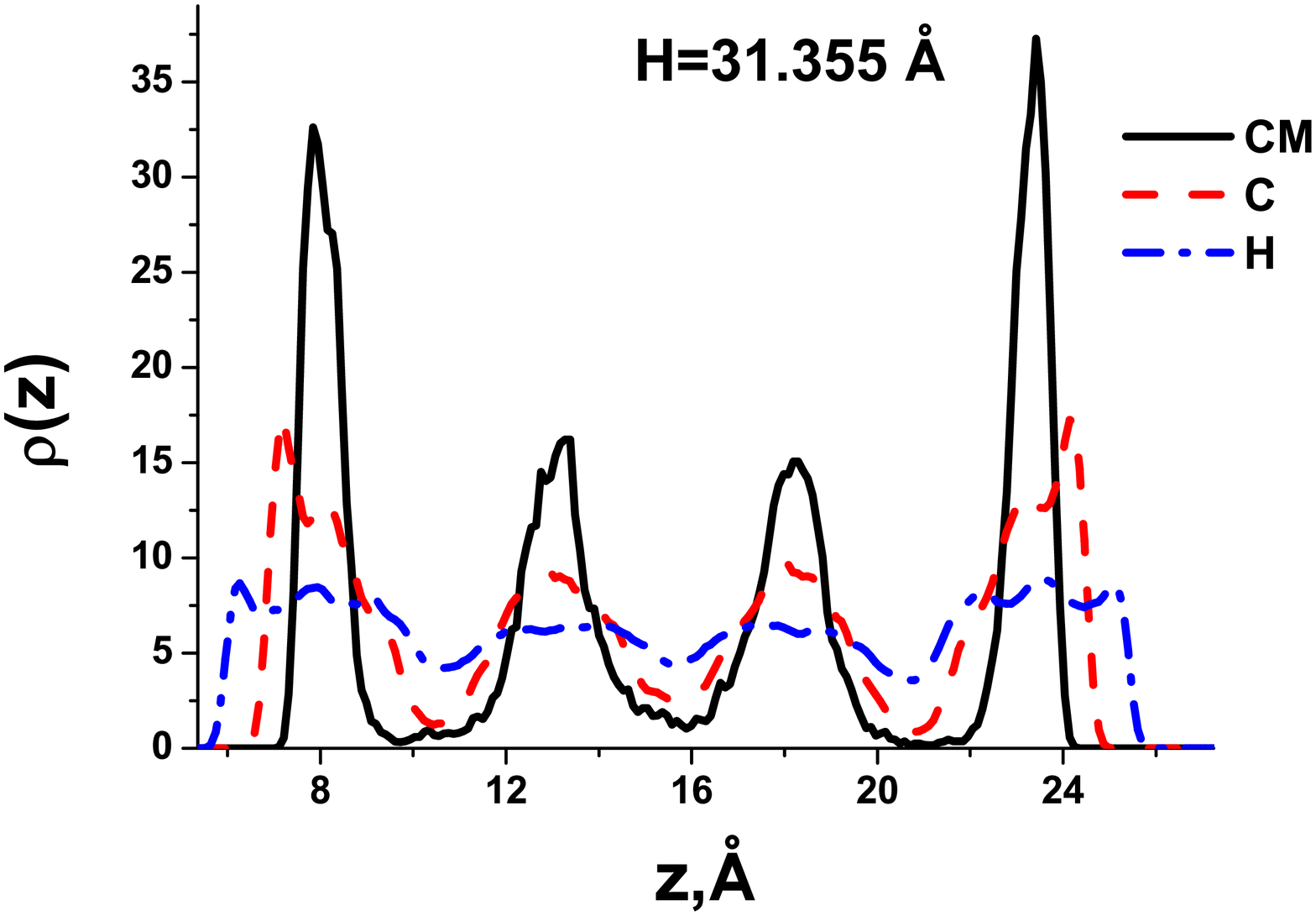}%
\includegraphics[width=5cm, height=5cm]{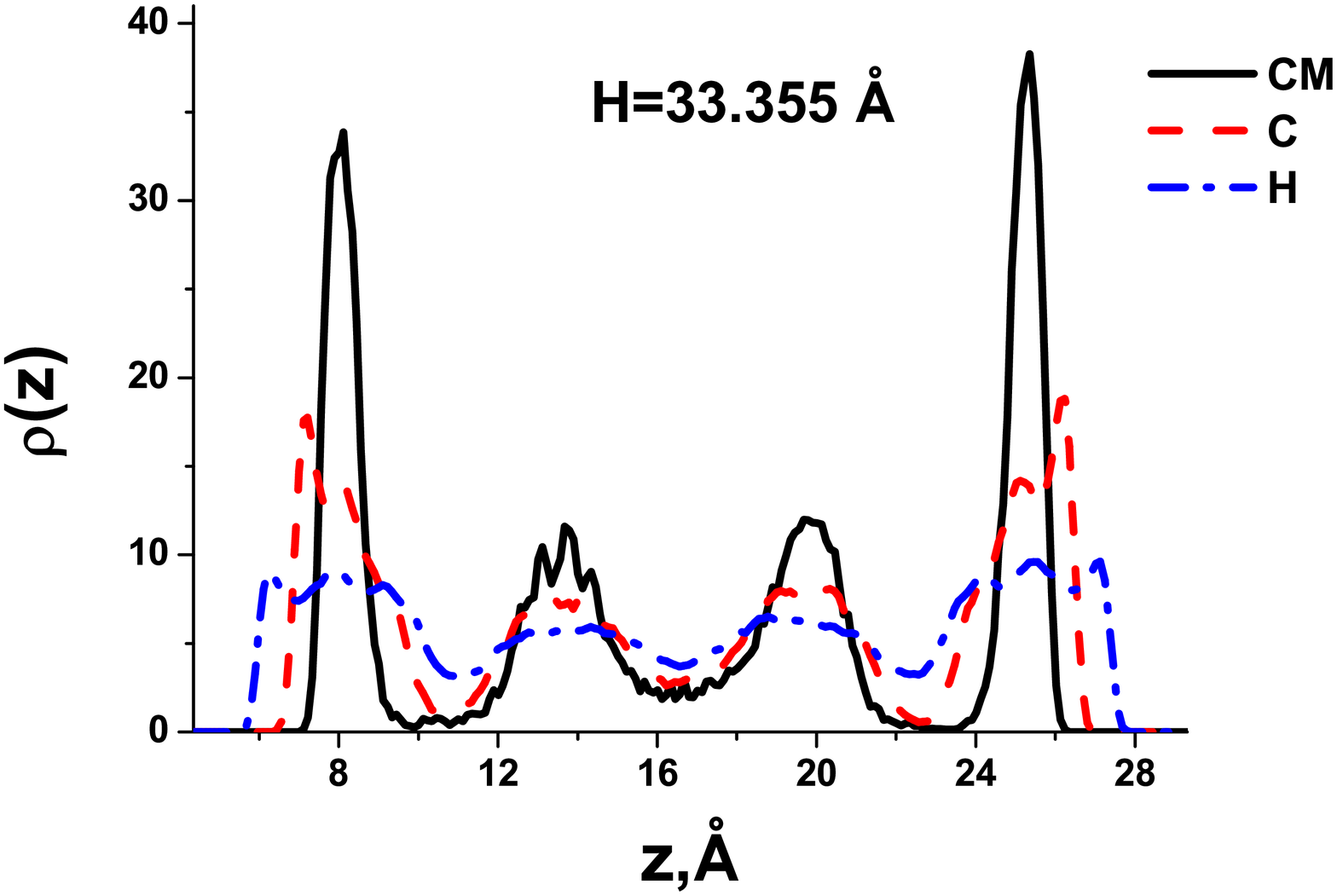}%

\caption{\label{fig:fig1} Distribution of number density of the
molecules centers of mass (CM), carbon (C) and hydrogen (H) atoms
perpendicular to the graphite wall for several pore sizes. In
order to have the same scale we multiply the CM curves by $6$ and
divide the H curves by $2$. The sizes of the pores are shown in
the corresponding plots.}
\end{figure}

As many other liquids, cyclohexane in a slit pore forms layered
structures. The number densities of carbon, hydrogen and centers
of mass of the molecules in the case of different pore heights are
shown in Fig.~\ref{fig:fig1}. One can see that in the case of
small pores ($H \leq 27.355 \AA$) three layers are formed: two
next to the lower and upper walls and one in the center, while at
larger separation the peak in the center splits into two ones. It
is clearly seen from the density distribution of carbons and
centers of mass. At the same time the density of hydrogen atoms
looks almost flat.

Below we call the layers close to the upper and lower walls as the
outer ones while the layer or two layers in the center of the pore
as the inner ones.

\begin{figure}
\includegraphics[width=5cm, height=5cm]{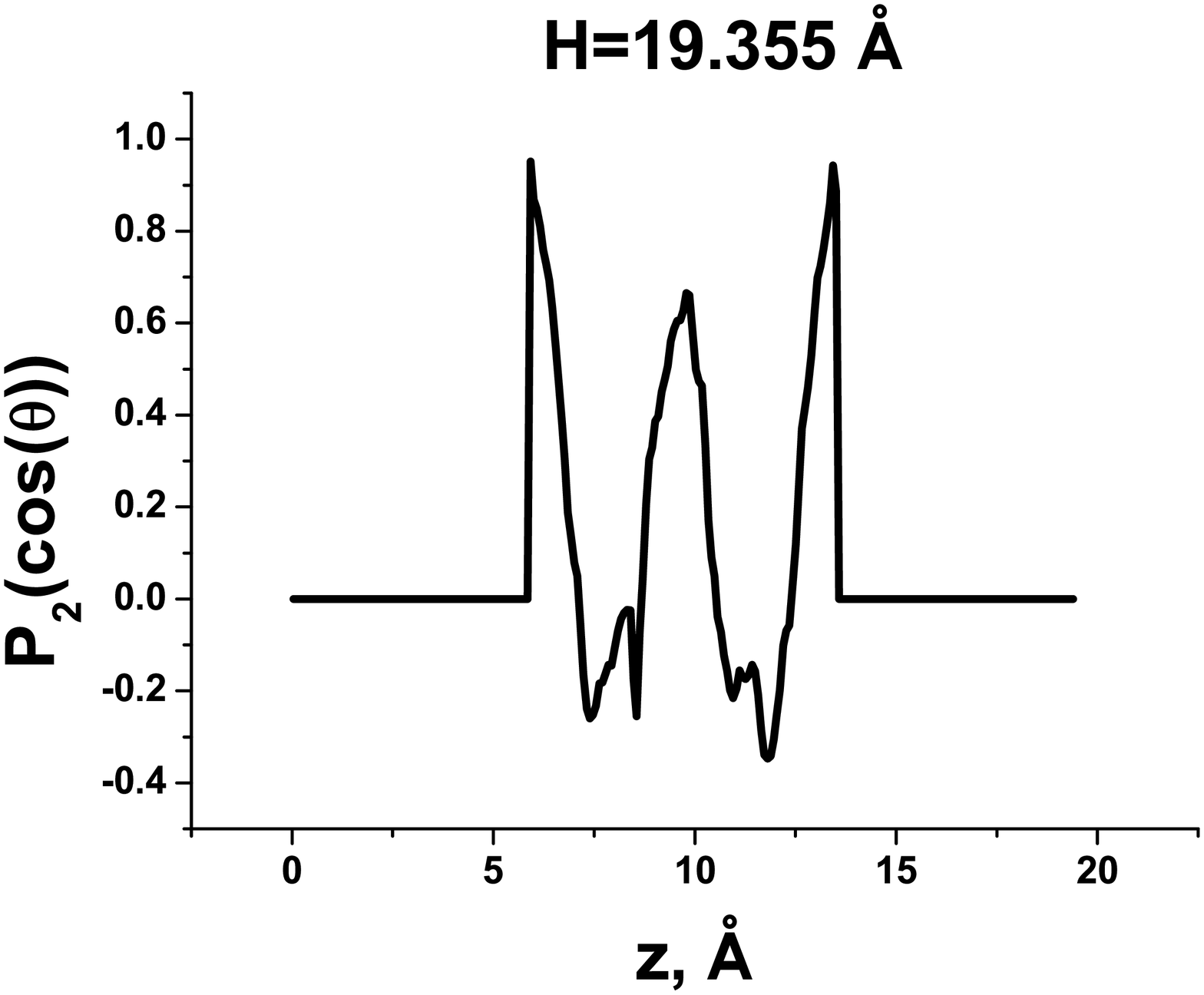}%
\includegraphics[width=5cm, height=5cm]{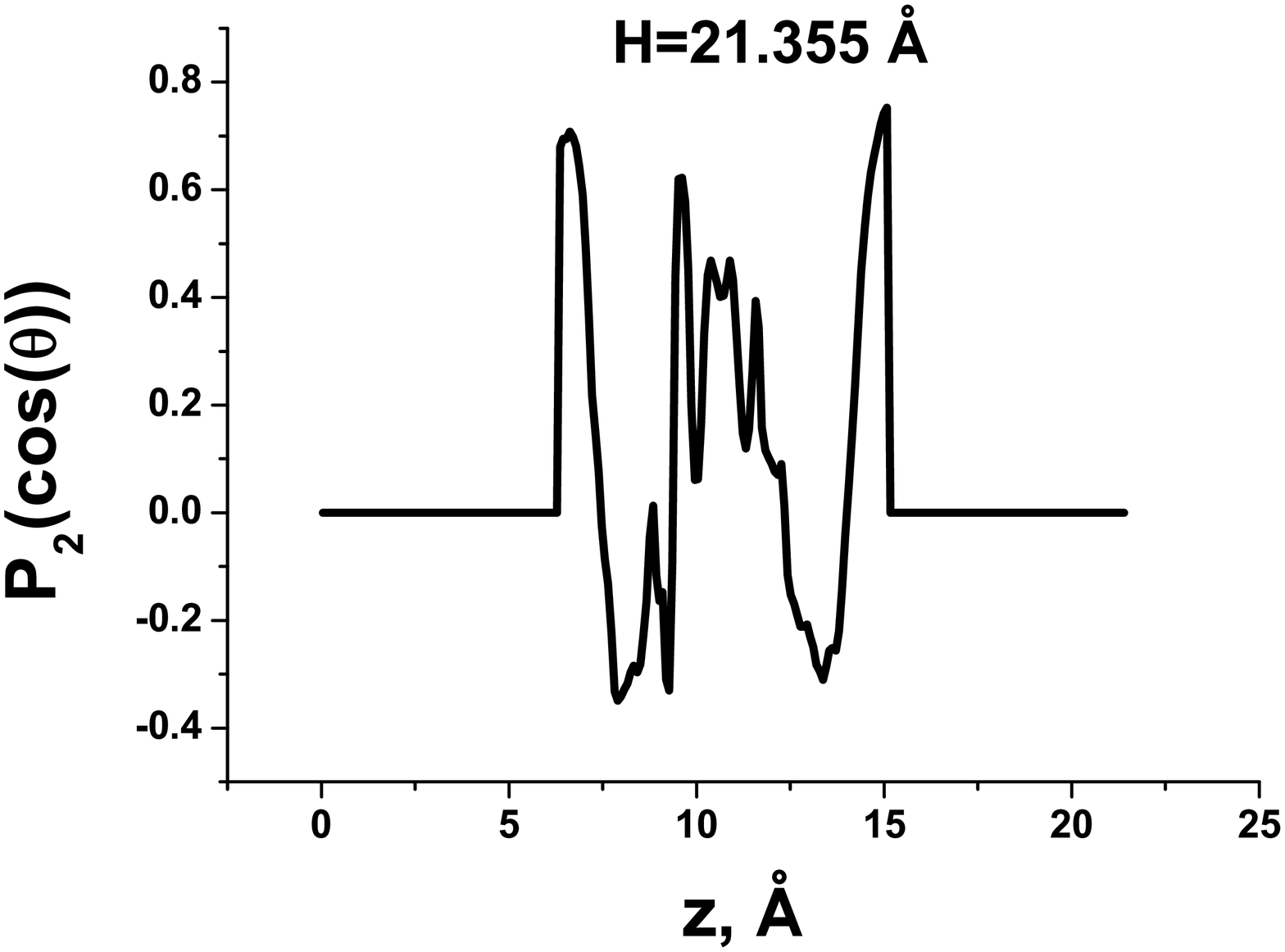}%

\includegraphics[width=5cm, height=5cm]{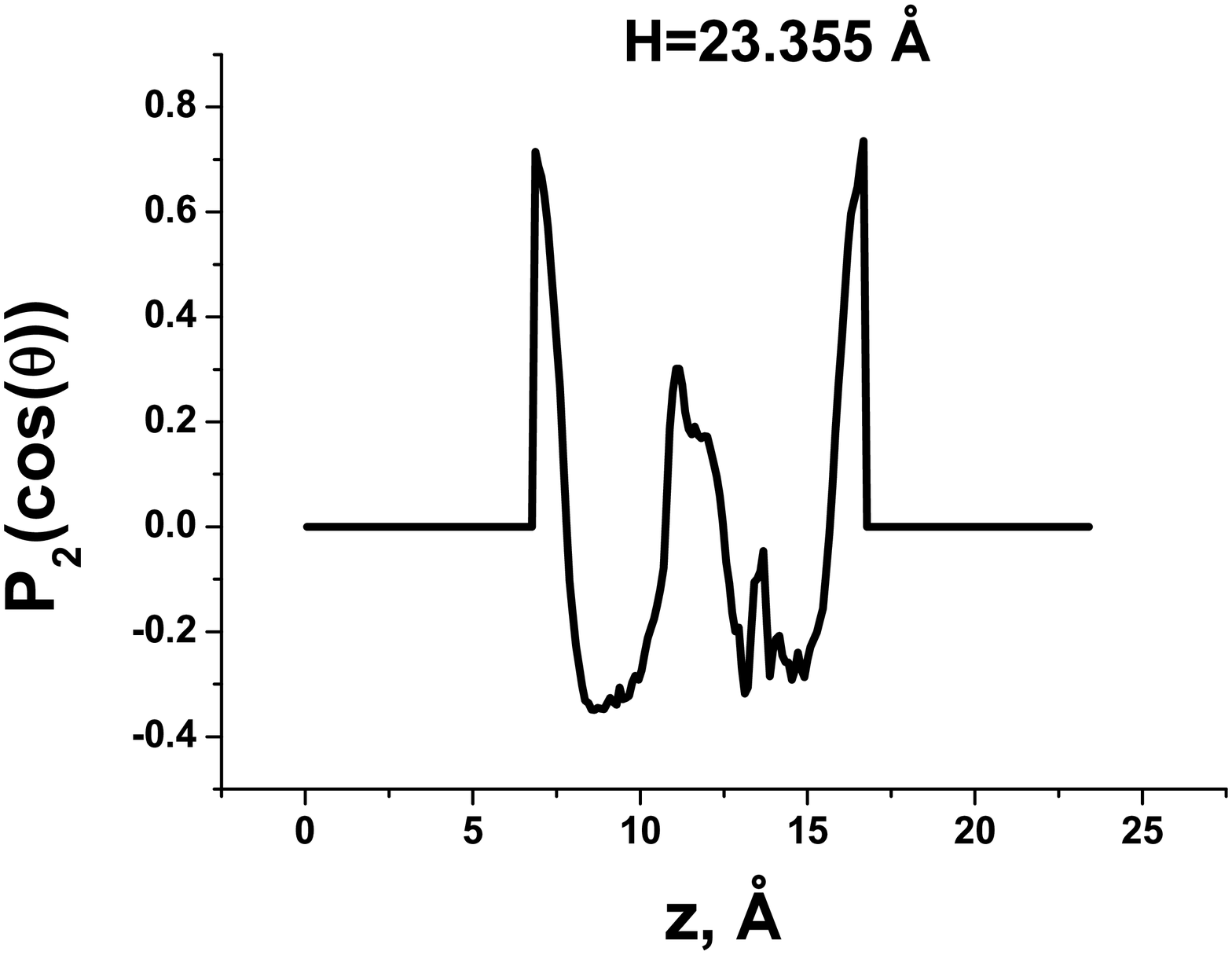}%
\includegraphics[width=5cm, height=5cm]{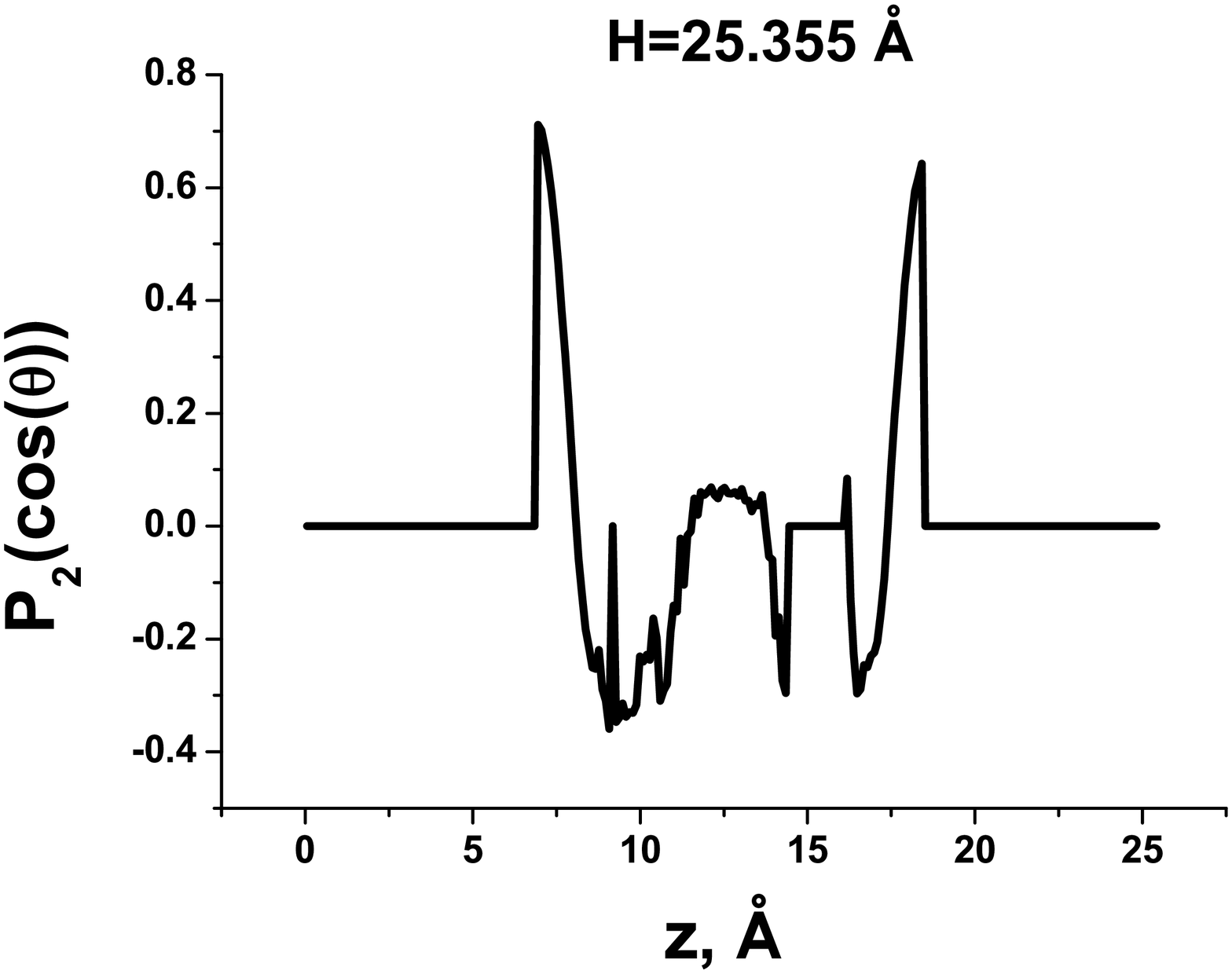}%

\includegraphics[width=5cm, height=5cm]{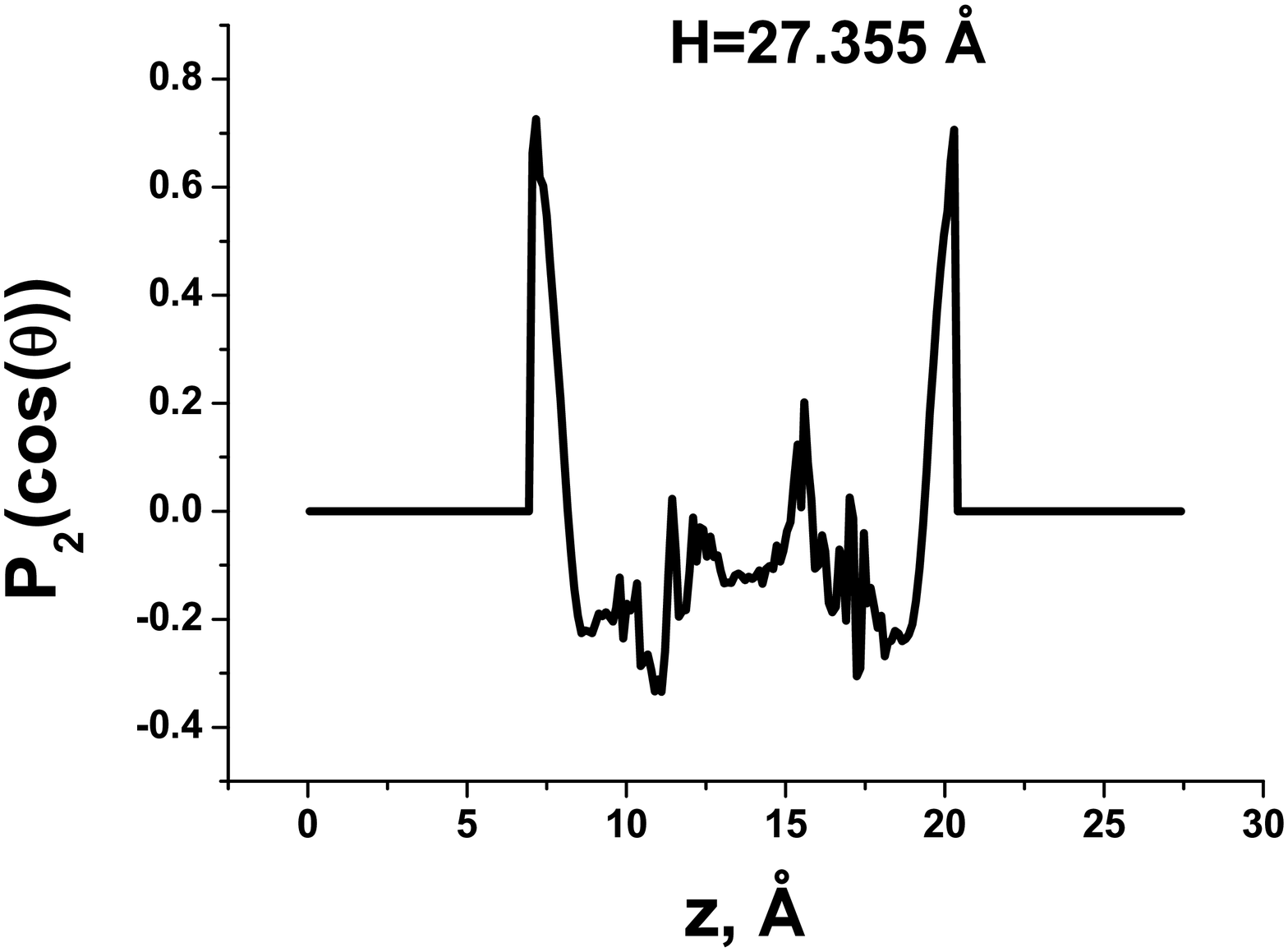}%
\includegraphics[width=5cm, height=5cm]{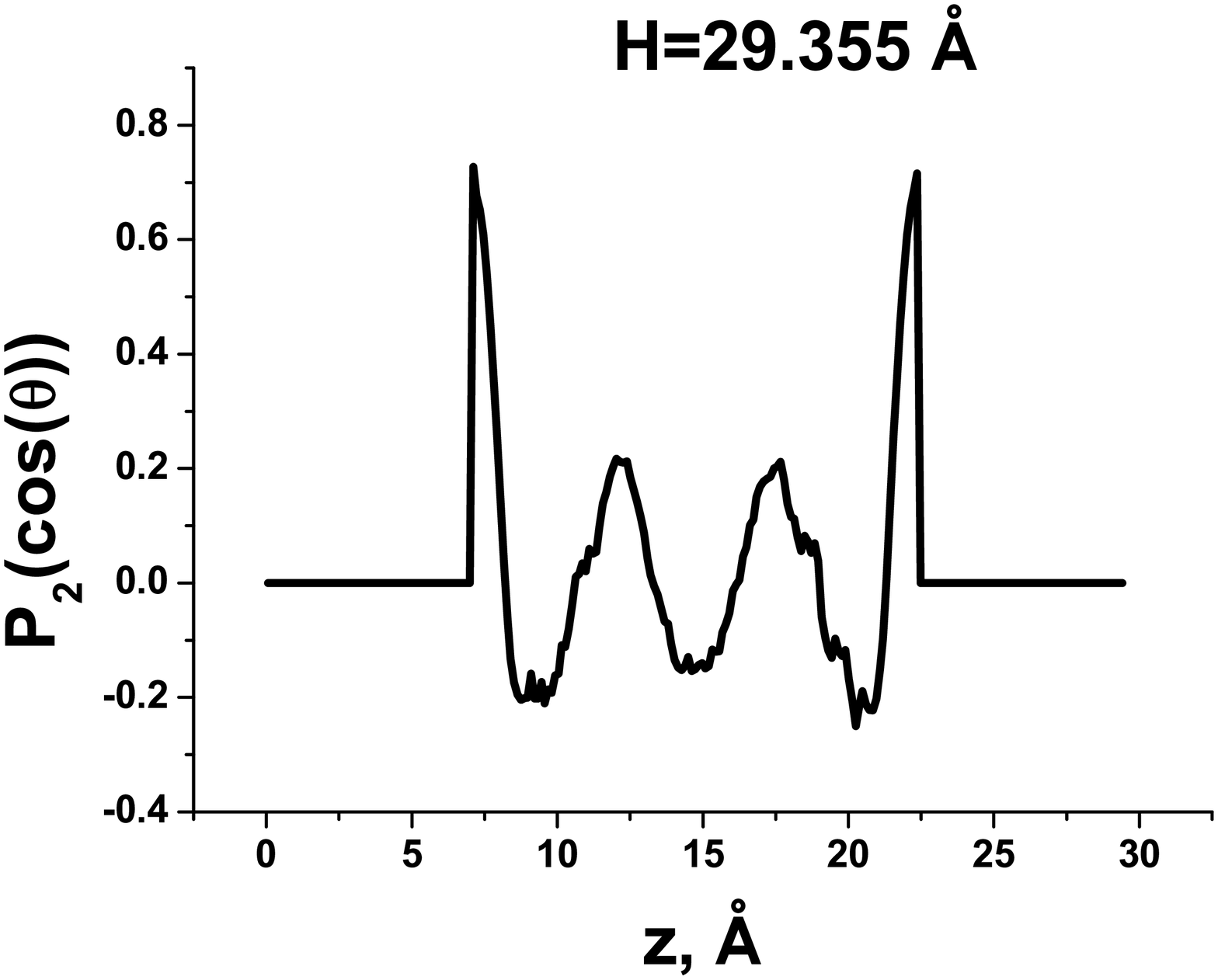}%

\includegraphics[width=5cm, height=5cm]{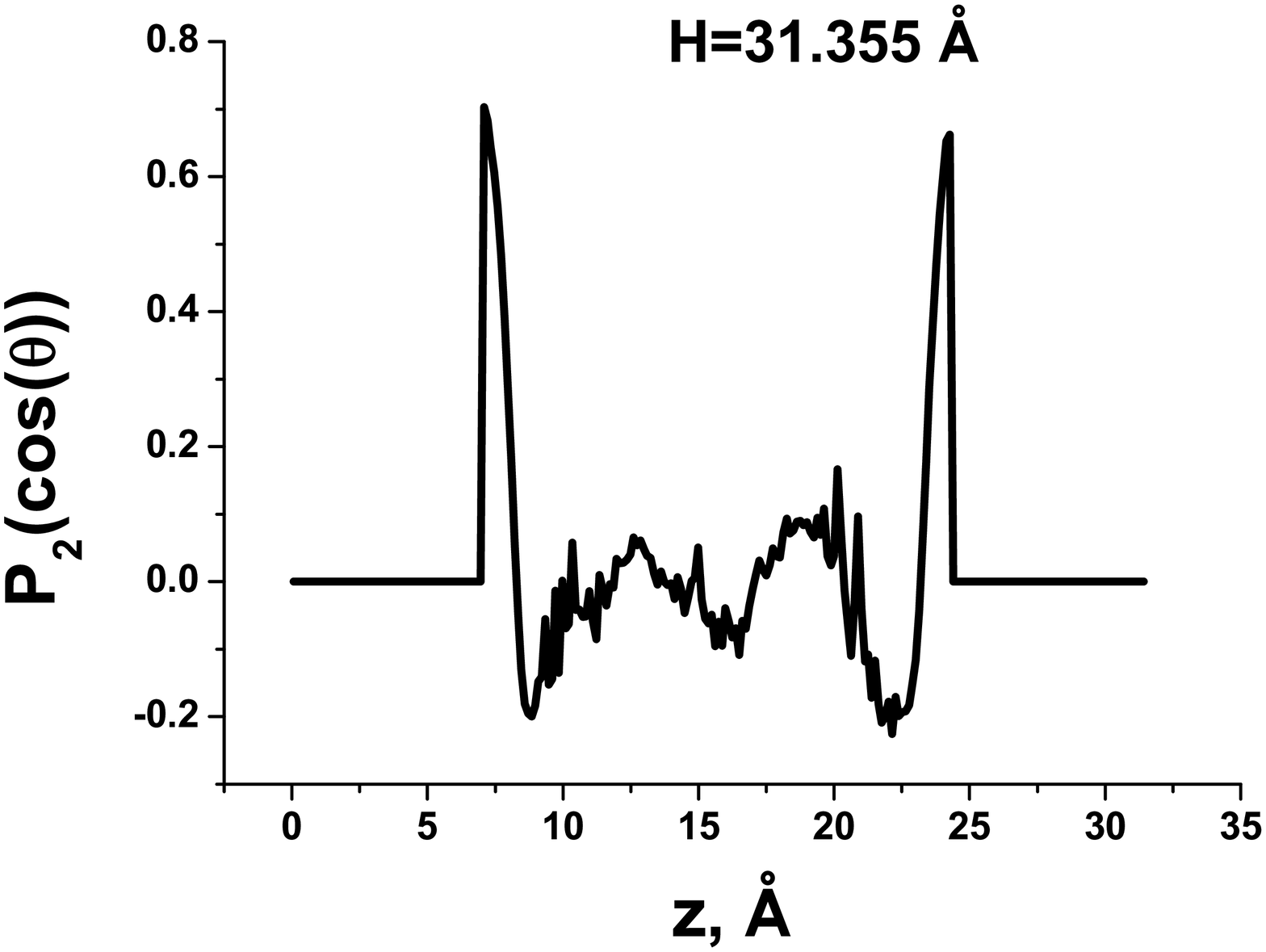}%
\includegraphics[width=5cm, height=5cm]{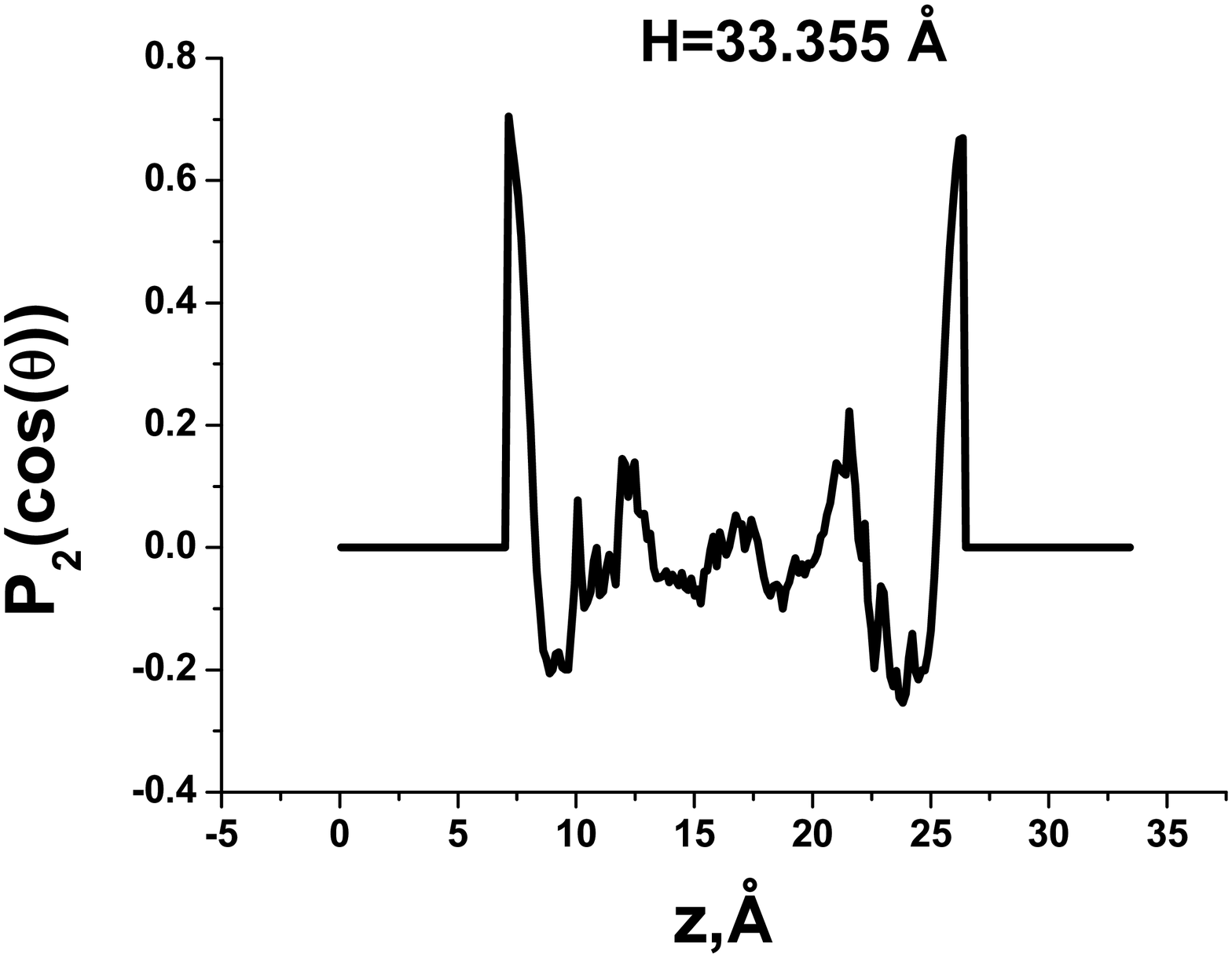}%

\caption{\label{fig:fig2} The distribution of second order
Legender polynomial $P_2(cos(\theta))$ along $z$ axis of the
system for different pore sizes.}
\end{figure}

The density distribution of the carbon atoms is very sensitive to
the pore height. One can see that at the height as small as
$21.355 \AA$ a shoulder on the outer peaks appears. Later on it
transforms into a small subpeak. At $H=27.355 \AA$ the inner peak
splits into two subpeaks. On increasing $H$ these subpeaks become
separate peaks.

The density distribution of CM are less sensitive to the change of
$H$. One can not see small subpeaks and splitting at $H=27.355
\AA$. However, one observes that the central peak splits into two
well defined peaks on change from $H=29.355 \AA$ to $H=33.355
\AA$.

Next we consider the orientation of the cyclohexane molecules
inside the pore. It is common to characterize the orientation of
molecules by second order Legender polynomial
$P_2(cos(\theta))=1.5 cos^2(\theta)-0.5$, where $\theta$ is the
angle between some vector which can determine the molecule
orientation (for example, in the case of benzene it is the normal
vector to the plane of the benzene ring) and $z$ axis.
$P_2(cos(\theta))=1$ means that the molecule is parallel to the
wall, while $P_2(cos(\theta))=-0.5$ corresponds to perpendicular
orientation of the molecule. In our previous publication we
considered benzene confined in graphite pore \cite{benzen-me}. It
was shown that in the case of small pores the molecules lie
parallel to the walls forming a planar structure. However, it can
be related to the planar shape of the benzene molecules. Unlike
the case of benzene the molecules of cyclohexane are not planar,
which makes the situation more complex. First of all, in the case
of benzene one characterize the orientation of the molecule by the
normal vector to the benzene ring plane. The carbon ring of
cyclohexane molecule is not planar so one needs to adjust the
definition.

A procedure for non-planar molecules was proposed in our previous
publication devoted to the benzene inside a carbon nanotube
\cite{benzene-nanotube}. For the sake of completeness we give a
brief description of this procedure below.

Let us denote all carbon atoms in the ring by numbers from $1$ to
$6$ and define the vectors connecting these atoms: $\bf {r}_{12}$,
$\bf {r}_{13}$ and so on. Each pair of such vectors can be used to
define a plane. The normal vector to this plane can be calculated
as the vector product of the corresponding vectors between the
atoms. Employing this procedure we can calculate $20$ vectors
perpendicular to the planes of triplets of the carbon atoms and
calculate $P_2(cos(\theta))$, where $\theta$ is the angle between
the $z$ axis and the normal vector of each plane. The final
orientational parameter $P_2$ is defined as an arithmetic average
of all $20$ values for different triplets.

The distributions of $P_2$ along the $z$ axis are shown in
Figs.~\ref{fig:fig2} (a)-(g). One can see that in all considered
cases the molecules at the outer layers lye almost parallel to the
graphite sheets. At $H=19.355 \AA$ the corresponding peaks are
almost unity. Upon increasing of $H$ these peaks slightly decay,
but even at $H=33.355 \AA$ they are still as high as $0.7$.

Unlike the outer peaks the inner one rapidly smashes out. One can
see that already at $H=25.355 \AA$ it practically disappears.
Later on it splits into two peaks, but both of them are small and
noisy which means that the orientational order is almost lost.

Importantly, in between of the layers the most preferable
orientation of the molecules is mainly perpendicular to the wall.
It means that the molecules located between the layers construct
some kind of bridges between them. This phenomena can be important
for stability of the layered structure.

As it was described above the system can be divided into layers.
These layers can be considered as quasi two dimensional systems.
This allows us to consider in-plane structure of each layer. The
structure of these two-dimensional layers can be characterized by
in-plane radial distribution functions $g_2(r)$.

In order to find the borders of the layers we took the minima of
the density distribution of the centers of mass
(Figs.~\ref{fig:fig1}). These borders are given in Table II. Note
that the width of the layers can be as large as $5-6 \AA$, i.e.
the layers are not perfectly flat. Therefore the two-dimensional
representation is not exact. However, since the width of the layer
does not exceed the size of the molecules quasi two-dimensional
consideration can be a useful approximate tool.

\begin{figure}
\includegraphics[width=5cm, height=5cm]{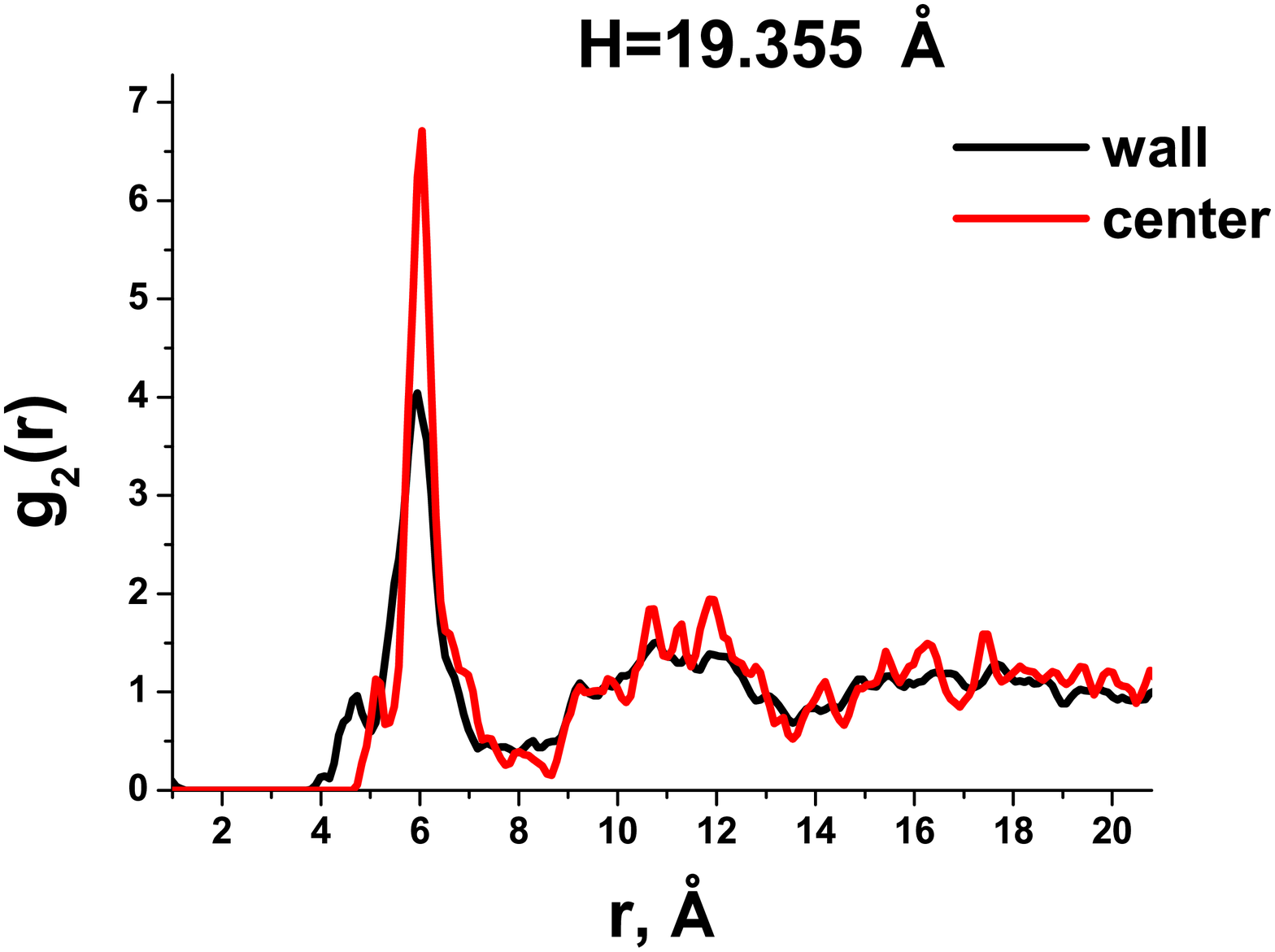}%
\includegraphics[width=5cm, height=5cm]{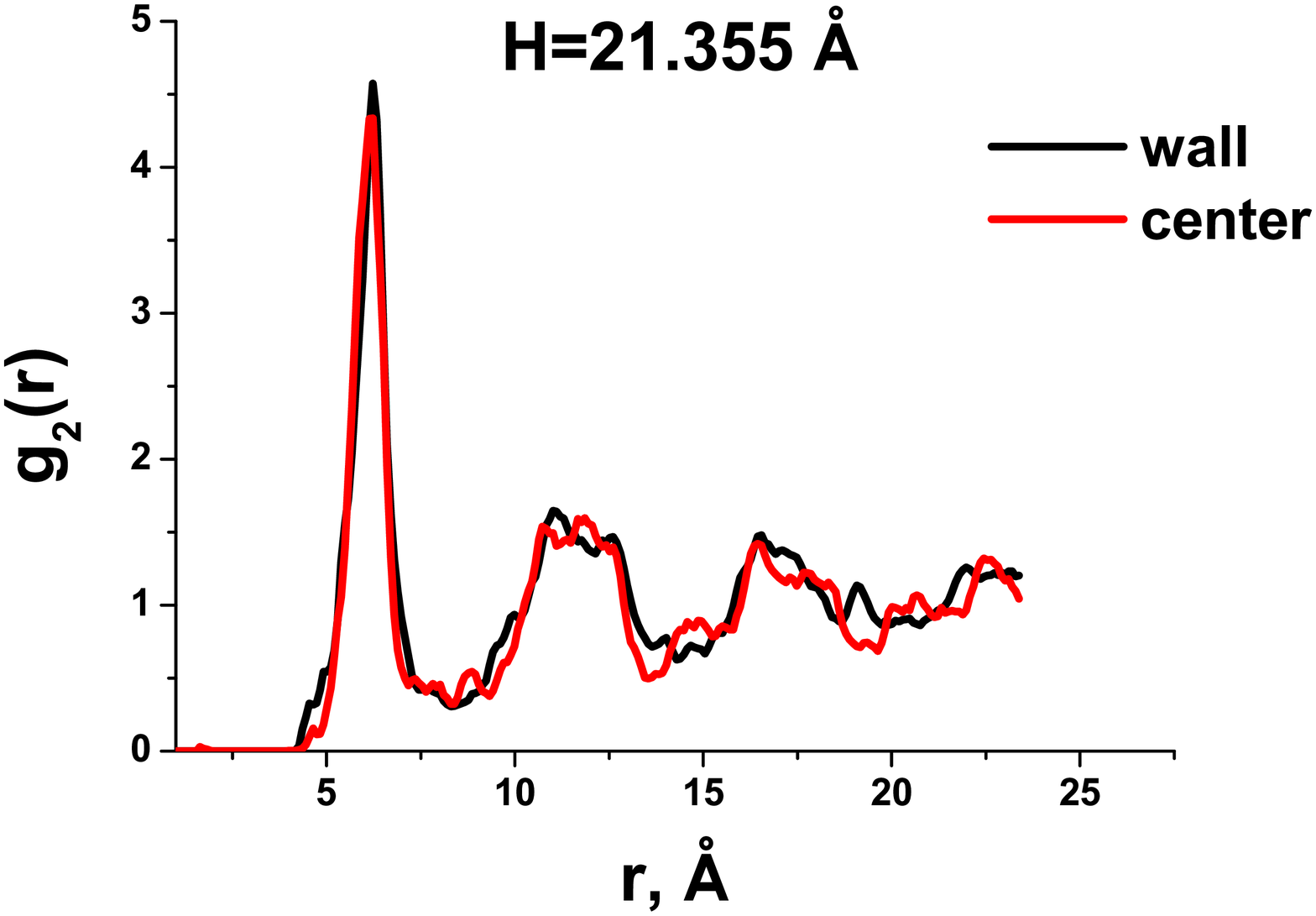}%

\includegraphics[width=5cm, height=5cm]{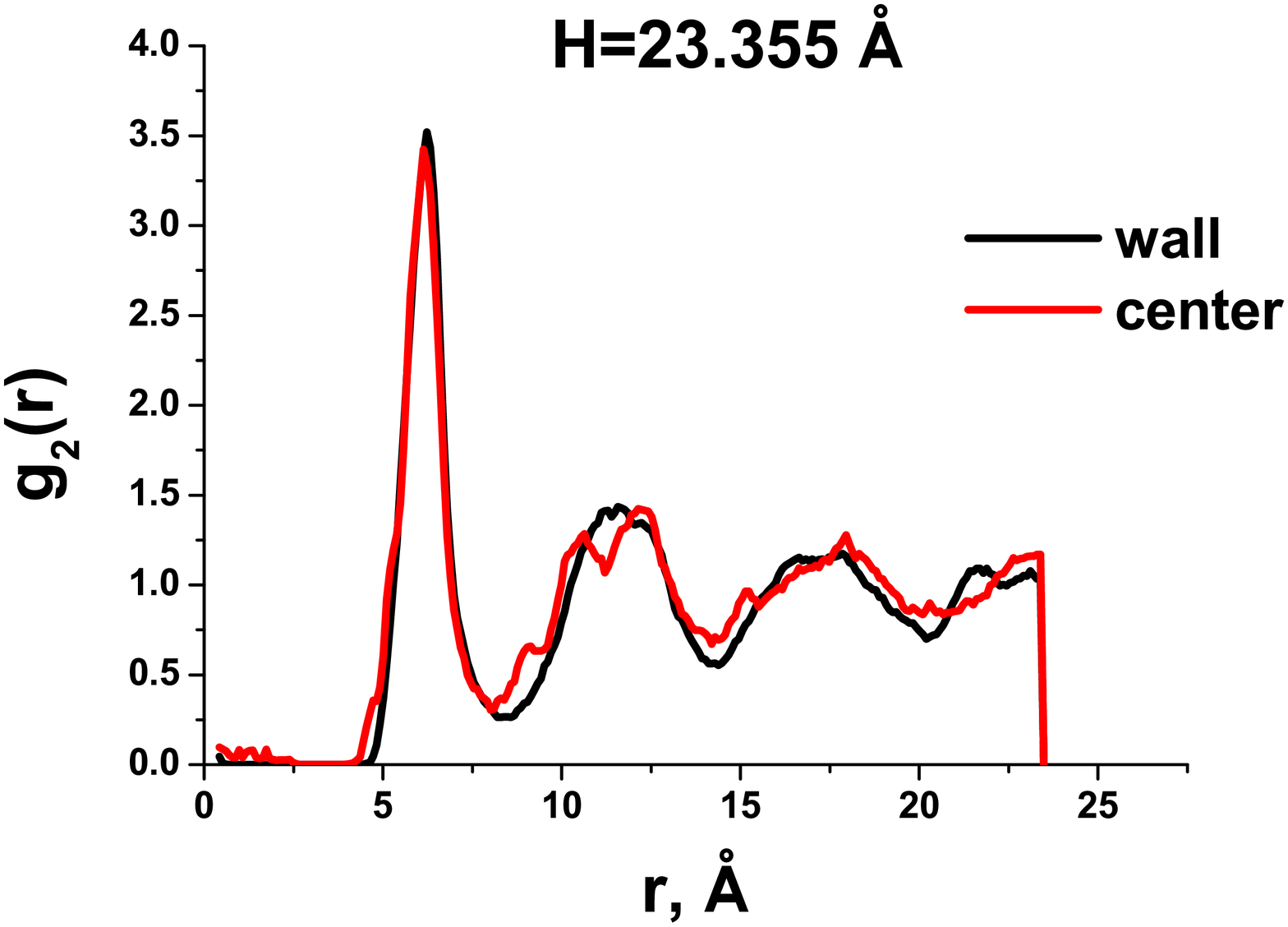}%
\includegraphics[width=5cm, height=5cm]{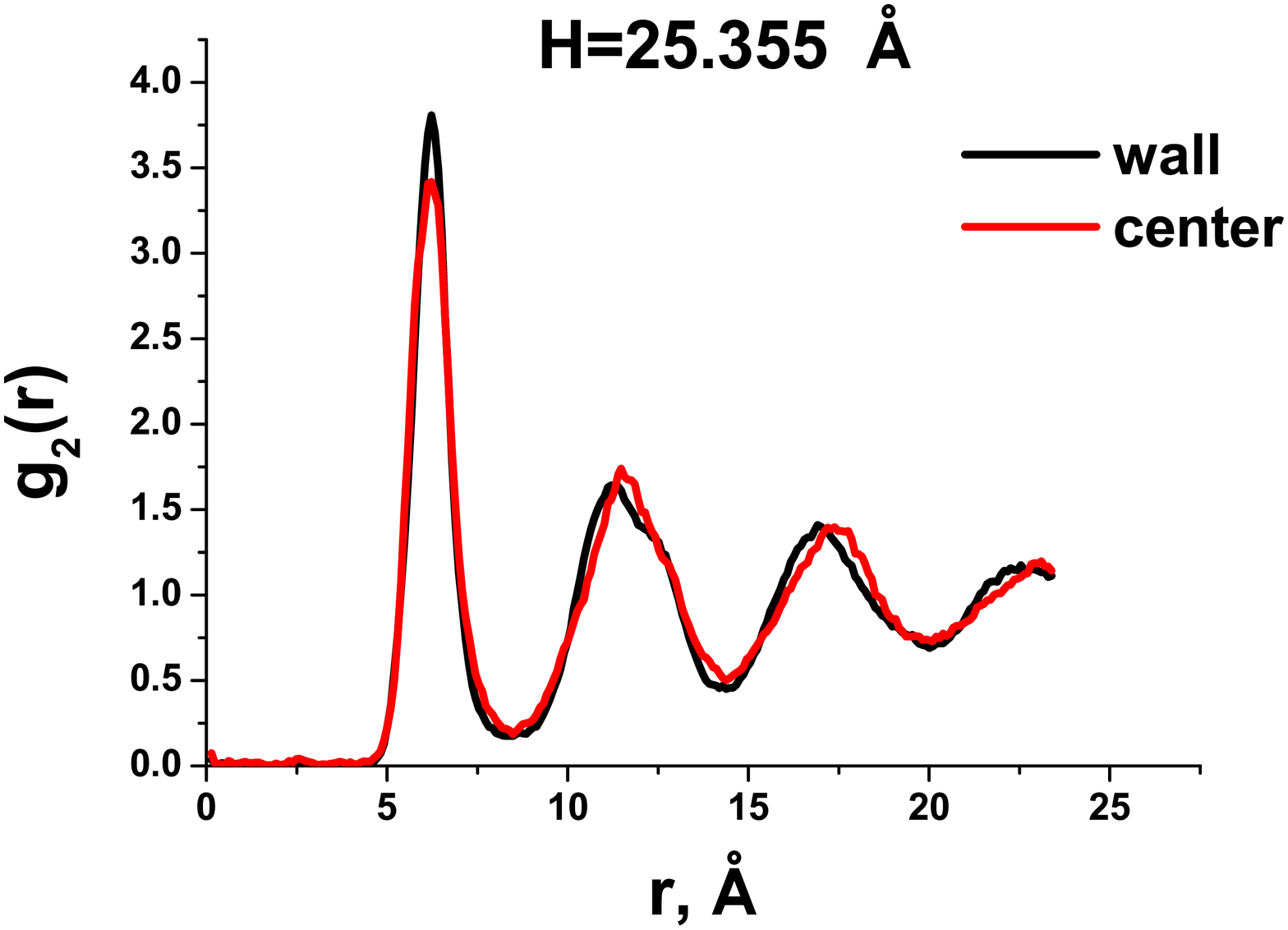}%

\includegraphics[width=5cm, height=5cm]{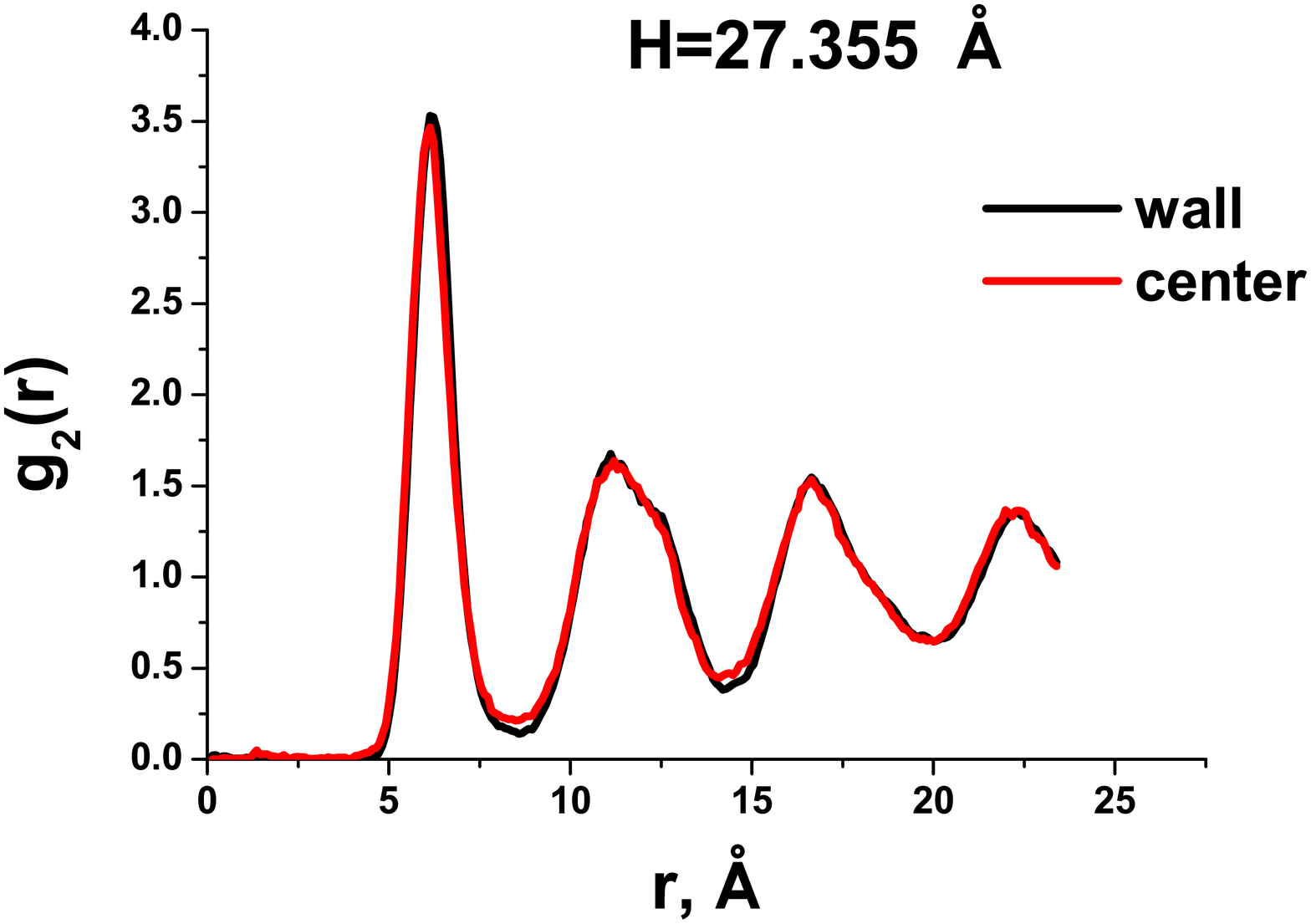}%
\includegraphics[width=5cm, height=5cm]{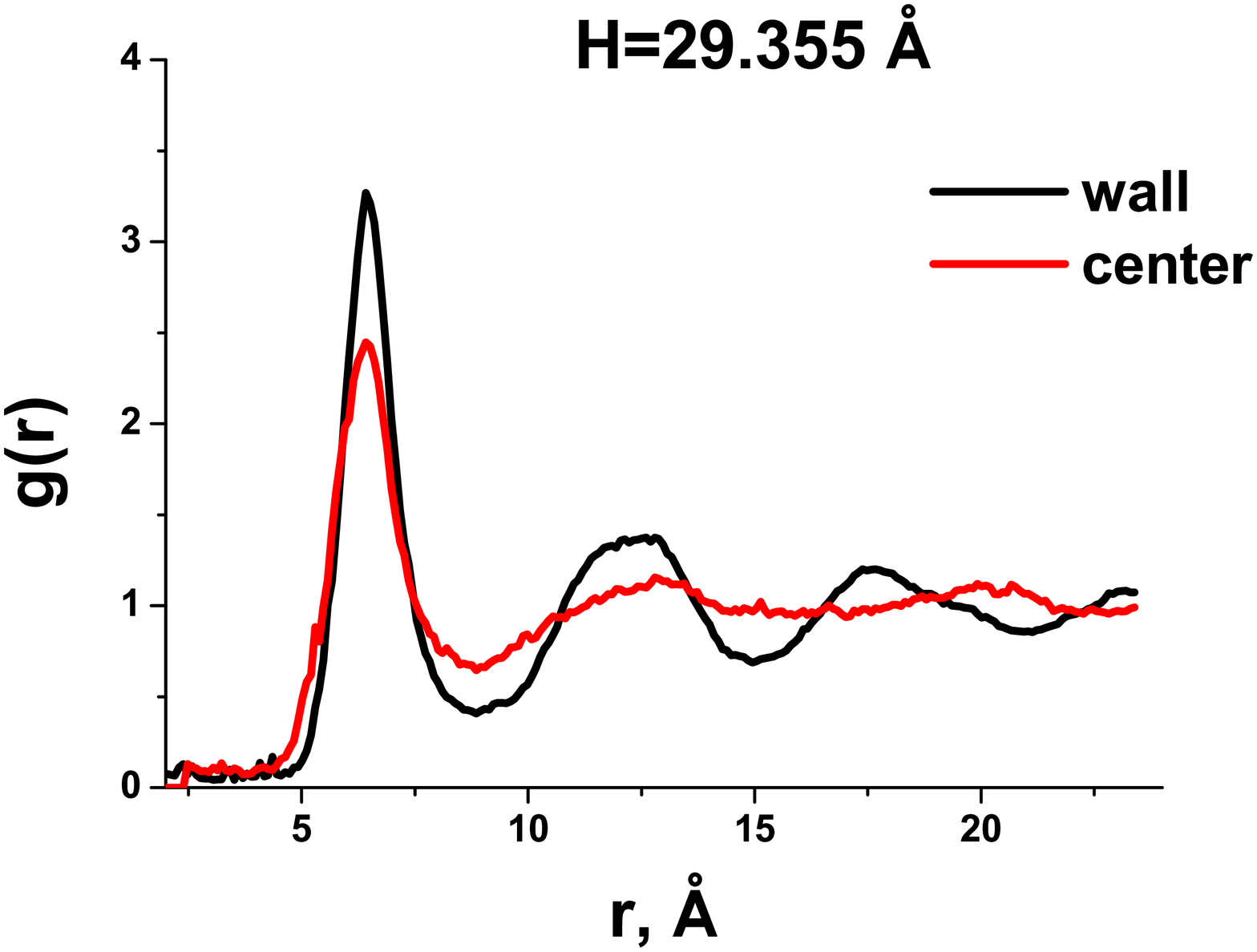}%

\includegraphics[width=5cm, height=5cm]{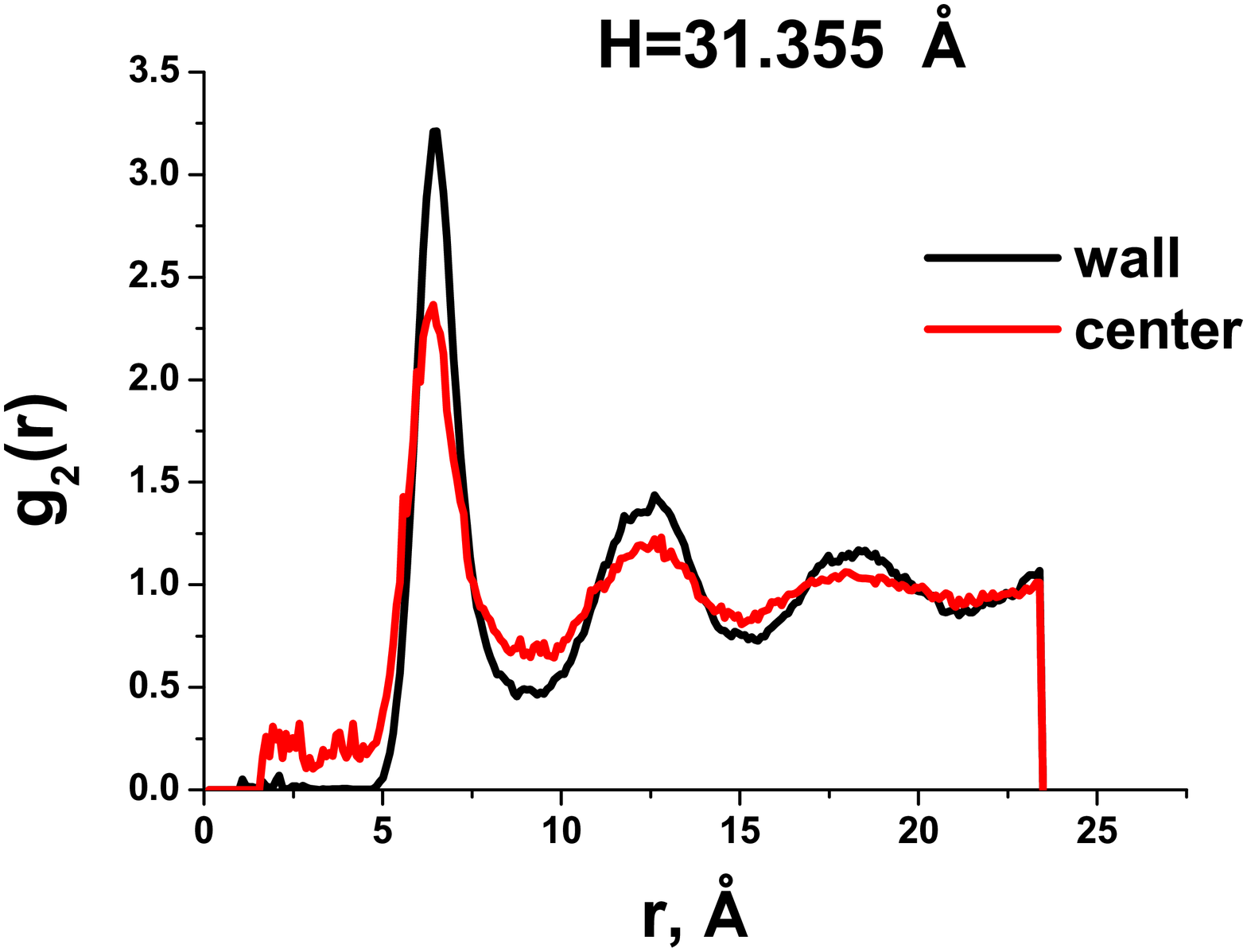}%
\includegraphics[width=5cm, height=5cm]{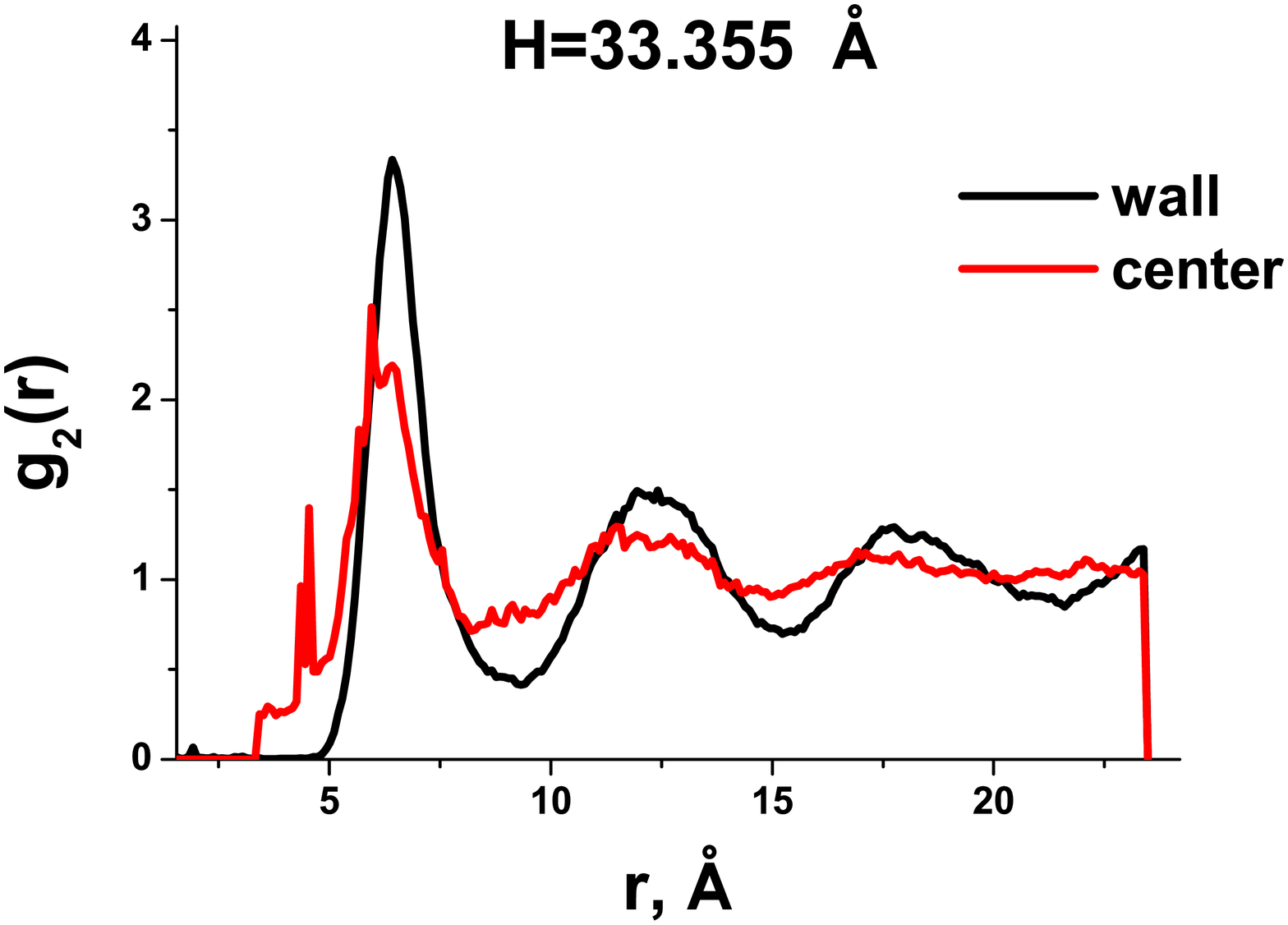}%

\caption{\label{fig:fig3} In-plane radial distribution functions
of centers of mass of $C_6H_{12}$ molecules in graphite pores. The
data are averaged for two peaks next to the wall ('wall' curve in
the plots). In the case of the peaks at the center at $H \geq
29.355 \AA$ the curve is averaged over two central peaks.}
\end{figure}

\begin{figure}
\includegraphics[width=5cm, height=5cm]{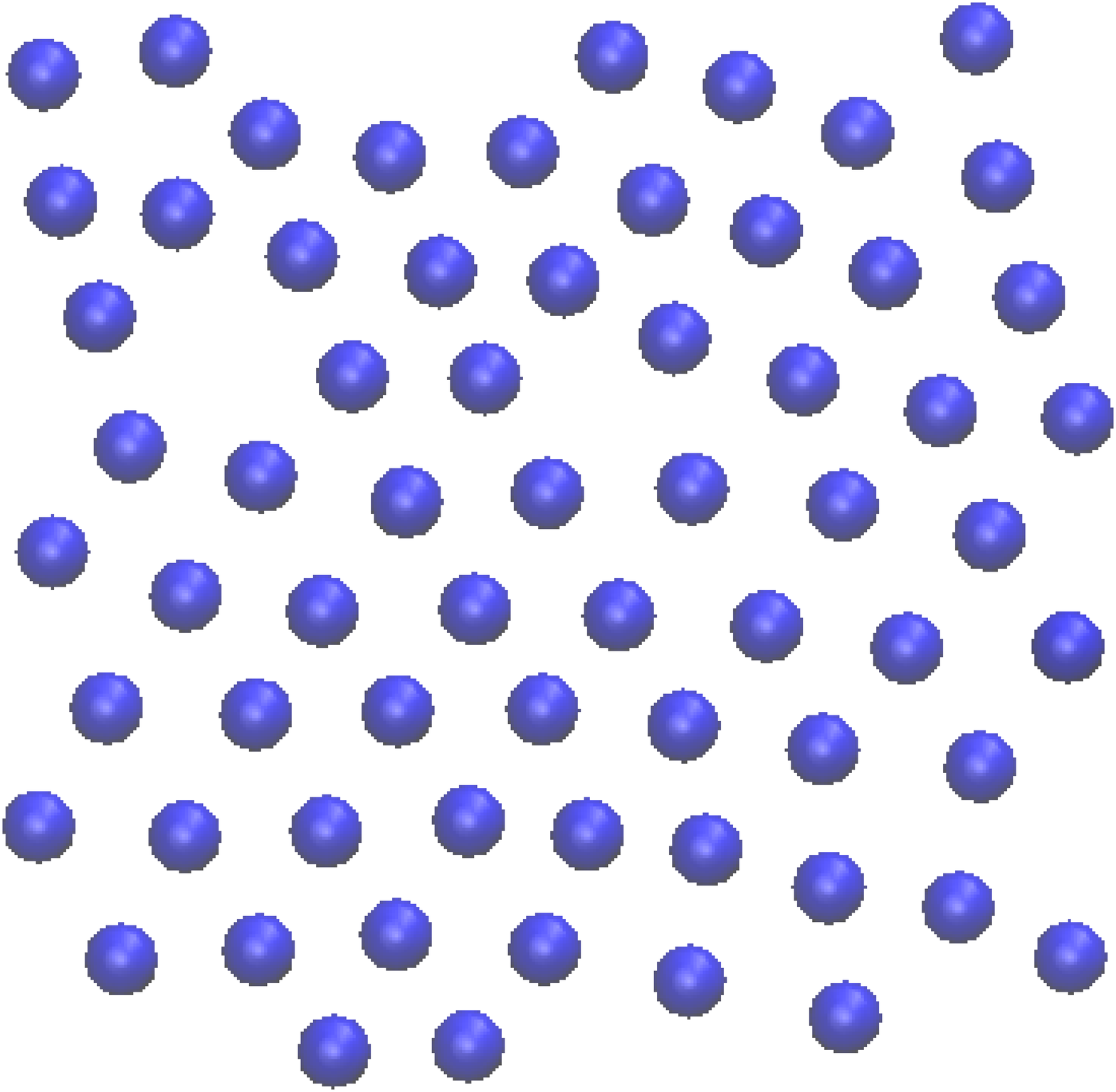}%
\includegraphics[width=5cm, height=5cm]{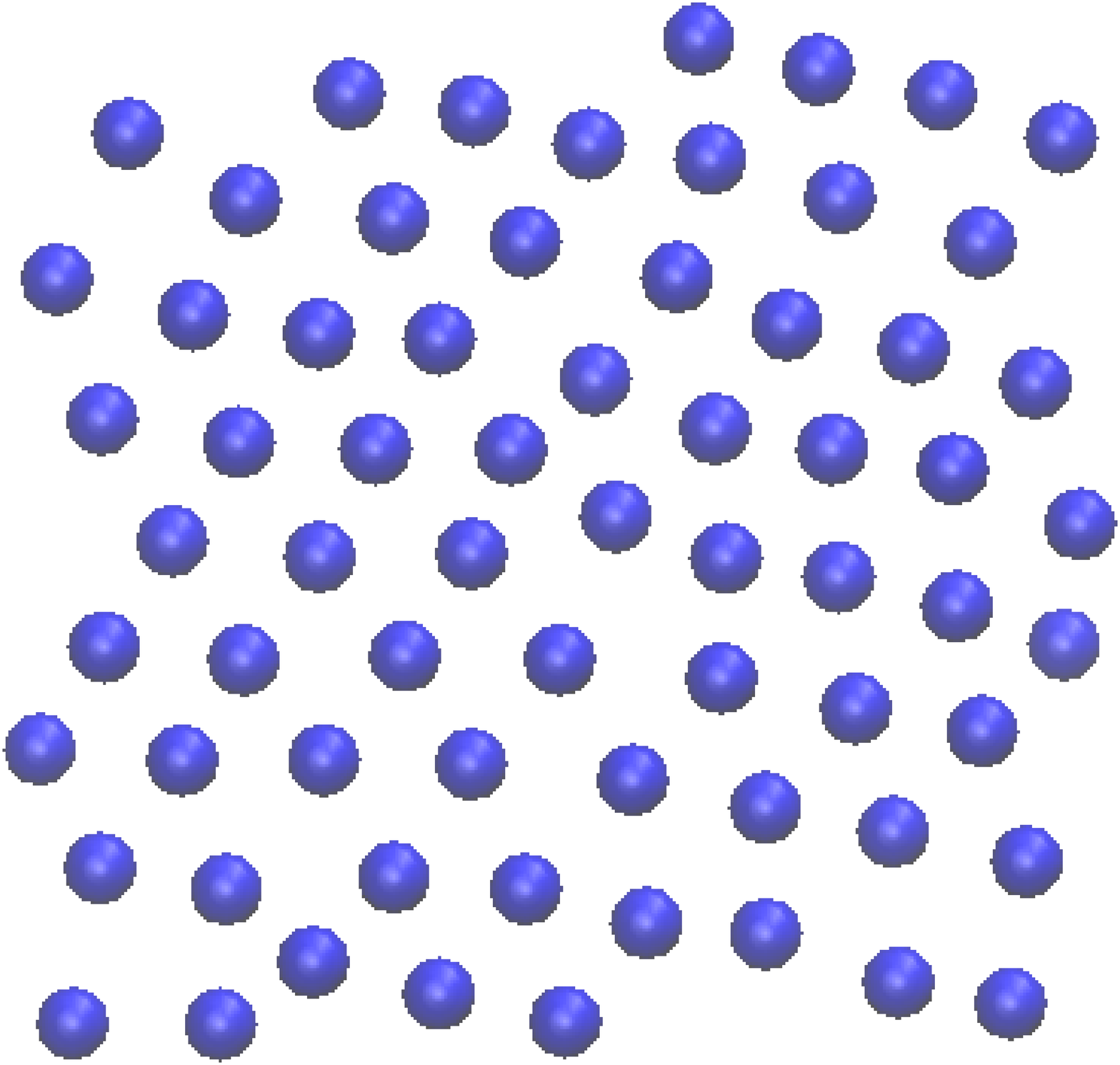}%

\caption{\label{fig:fig4} Left panel: Snapshot of the outer layer
of centers of mass of $C_6H_{12}$ molecules in $H=25.355 \AA$ slit
pore. Right panel: the same for the inner layer.}
\end{figure}

\begin{figure}
\includegraphics[width=5cm, height=5cm]{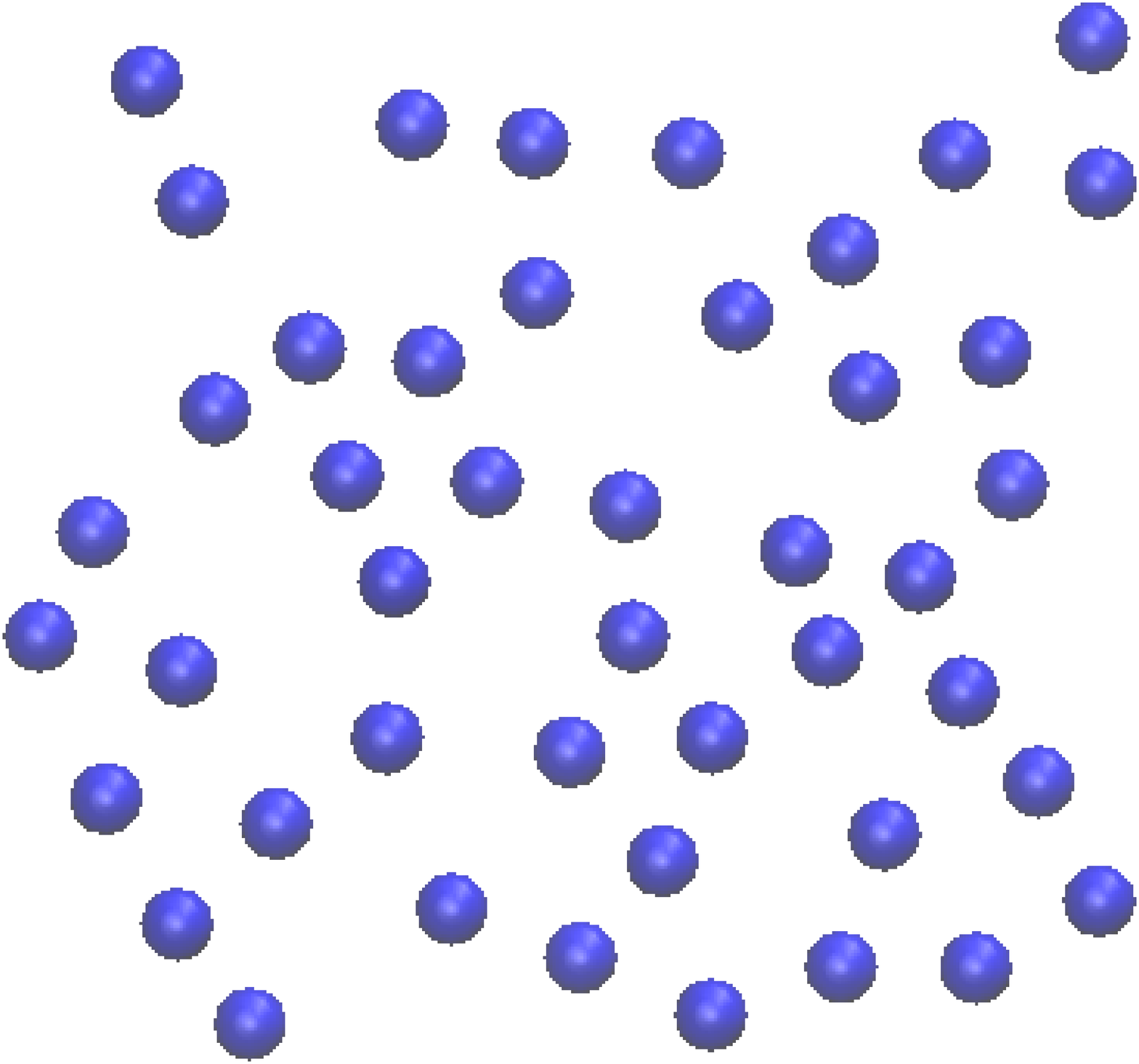}%
\includegraphics[width=5cm, height=5cm]{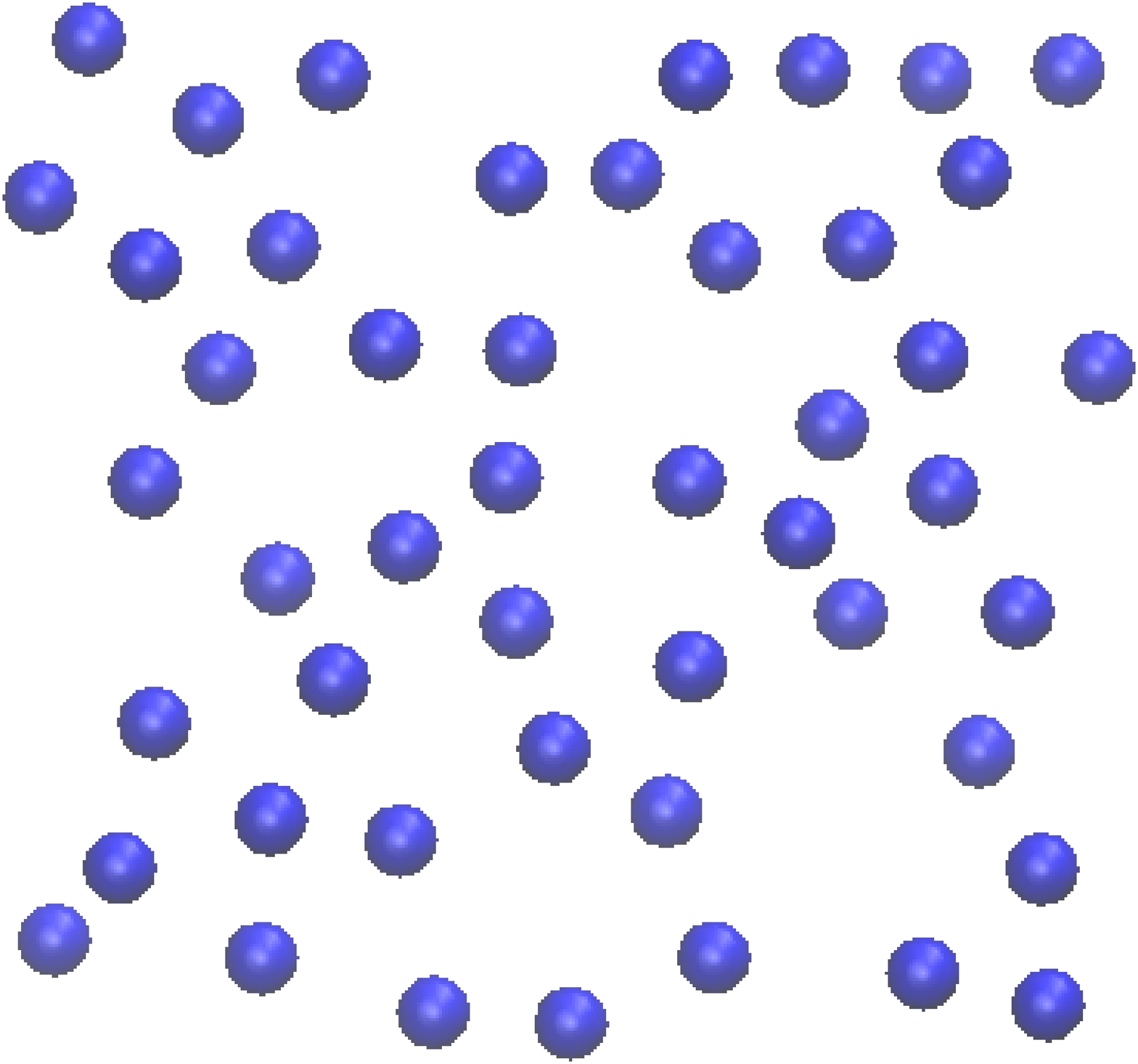}%

\caption{\label{fig:fig5} Left panel: Snapshot of the outer layer
of centers of mass of $C_6H_{12}$ molecules in $H=33.355 \AA$ slit
pore. Right panel: the same for the inner layer.}
\end{figure}

Two-dimensional radial distribution functions of centers of mass
of the molecules in different layers are shown in
Figs.~\ref{fig:fig3} (a)-(h). The results for outer peaks of
$\rho(z)$ (Fig.~\ref{figfig3})(curve 'wall') and for inner ones
(curve 'center') are shown. For the outer peaks of $\rho(z)$ the
radial distribution function $g_2(r)$ is averaged over the upper
and the lower layers. The same averaging is applied to the inner
layers if the central peak splits into two ones. One can see that
at $H=19.355 \AA$ the first peak of $g_2(r)$ for the central layer
is much larger then for the outer ones. We suppose that at such
low walls separation the dynamics of liquid is too slow and the
results strongly depend on the initial configuration. The
considerable splitting of the second peak of the radial
distribution function allows to suppose that in this case there is
a glassy state of the system, however, more thorough investigation
of this point is necessary.
%The results from
%$H=19.355 \AA$ up to $H=23.355 \AA$ can be discarded for this
%reason.
For $H=25.355 \AA$ one can observe $4$ peaks of $g_2$ which
corresponds to very structured liquid. Moreover the positions of
the peaks is (for $H=25.355 \AA$) $r_1=6.23 \AA$, $r_2=11.1 \AA$,
$r_3=16.9 \AA$ and $r_4=22.26 \AA$. The ratio of $r_2/r_1$ and
$r_3/r_1$ corresponds to hexagonal two-dimensional lattice.
Fig.~\ref{fig:fig4} shows a snapshot of outer layer (left panel)
and inner layer (right panel) for this value of $H$. One can see
that in both cases defected triangular lattices are formed. One
can conclude from both $g_2(r)$ for $H=25.355 \AA$
(Fig.~\ref{fig:fig3}) and the snapshots of the layers
(Fig.~\ref{fig:fig4}) that the inner and outer layers of this
system are structurally equivalent.

The same conclusion holds true for the system with $H=27.355 \AA$.
In the case of $H=29.355 \AA$, $H=31.355 \AA$, and $H=33.355 \AA$
the situation looks different. The crystallinity of both inner and
outer layers drops. This effect is especially noticeable in the
inner layers where only first peak of $g_2$ is well pronounced,
i.e. the centers of mass of the molecules in the inner layers
behave as moderate density liquid. $g_2(r)$ of outer layer looks
like in the case of high density liquid. Snapshots of these cases
are shown in Fig.~\ref{fig:fig5}.

\begin{figure}
\includegraphics[width=7cm, height=7cm]{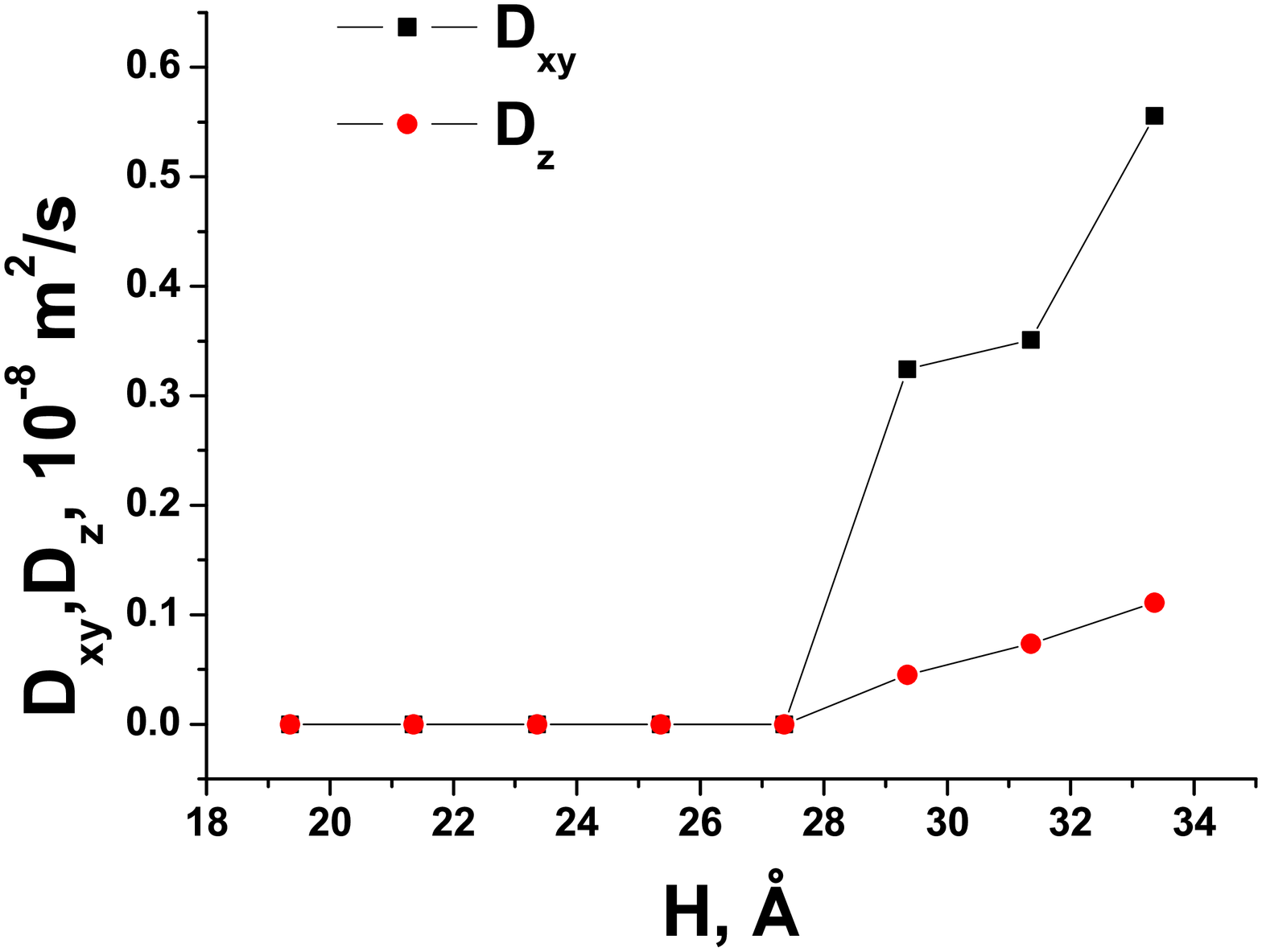}%

\caption{\label{fig:fig6} Transverse $D_z$ and longitudinal
$D_{xy}$ diffusion coefficients of the system as functions of $H$
at $T=300 K$.}
\end{figure}

To make the picture more clear, let us consider the behavior of
the transverse $D_z$ and longitudinal $D_{xy}$ diffusion
coefficients of the system as functions of $H$. $D_z$ and $D_{xy}$
are calculated with the help of the Einstein equation \cite{book}.
From Fig.~\ref{fig:fig6} one can see, that for $H\leq 27.355 \AA$
diffusion coefficient becomes equal to zero. Taking into account
Figs.~\ref{fig:fig4} and \ref{fig:fig5}, this behavior can be
interpreted as the freezing transition which occurs with
decreasing $H$. It should be noted that at $H=27.355 \AA$ the
number of layers changes from $4$ to $3$ (Fig.~\ref{fig:fig1}). It
is in qualitative agreement with the experimental results
\cite{mica01,mica2,mica02} on the behavior of cyclohexane in mica
slit pores. Moreover, in our case the freezing temperature is
$T=300 K$ whereas the bulk freezing temperature of cyclohexane is
equal to $T_b=279.62 K$. This increase of freezing temperature is
again in qualitative agreement with the experiment
\cite{mica01,mica2,mica02} and computer simulations of model
systems \cite{cui,ravi}.

\begin{table}
\begin{tabular}{|c|c|c|}
  \hline
  $H$, $\AA$ & $h$, $\AA$ & $\rho$, $g/cm^3$ \\
  \hline
  $17.172$ & $10.462$ & $1.075$ \\
  $18.172$ & $11.462$ & $0.982$ \\
  $19.172$ & $12.462$ & $0.903$ \\
  $20.172$ & $13.462$ & $0.836$ \\
  $21.172$ & $14.462$ & $0.778$ \\
  $22.172$ & $15.462$ & $0.728$ \\
  \hline
\end{tabular}

\caption{The size of the graphite pore $H$, the distance between
the inner graphite sheets $h$ and the density of confined benzene
$\rho$}
\end{table}

\begin{table}
\begin{tabular}{|c|c|c|c|}
  \hline
  H & Layer 1 & Layer2 & Layer 3 \\
  \hline
$H=17.172$ & $z_1=5.29$ & $z_1=7.97$ & $z_1=10.68$ \\
     & $z_2=6.49$ & $z_2=9.17$ & $z_2=11.71$ \\
  \hline
$H=18.172$ & $z_1=5.34$ & $z_1=8.21$ & $z_1=11.19$ \\
     & $z_2=6.61$ & $z_2=9.7$ & $z_2=12.68$ \\
  \hline
$H=19.172$ & $z_1=5.71$ & $z_1=8.78$ & $z_1=11.69$ \\
     & $z_2=7.01$ & $z_2=10.47$ & $z_2=13.61$ \\
  \hline
$H=20.172$ & $z_1=5.77$ & $z_1=8.96$ & $z_1=12.43$ \\
     & $z_2=7.63$ & $z_2=11.09$ & $z_2=14.48$ \\
  \hline
$H=21.172$ & $z_1=5.25$ & $z_1=9.14$ & $z_1=12.03$ \\
     & $z_2=9.14$ & $z_2=12.03$ & $z_2=15.16$ \\
  \hline
$H=22.172$ & $z_1=4.74$ & $z_1=9.13$ & $z_1=12.90$ \\
     & $z_2=9.13$ & $z_2=12.90$ & $z_2=16.81$ \\
  \hline
\end{tabular}

\caption{The coordinates of the borders of benzene layers for
different graphite pore size.}

\end{table}

%\section*{\sffamily \Large DISCUSSION}

%((Place Discussion here. Not needed for review articles.))

\section*{\sffamily \Large IV CONCLUSIONS}

%((Place Conclusions here.))
In conclusion, we present the molecular simulation study in NVT
ensemble of the cyclohexane confined between the carbon plates
(see Fig.~\ref{fig:fig0}) using the realistic AIREBO potential.
The local structure of the system is studied as a function of the
distance between the plates $H$ and it is shown that the molecules
form the layered structure. The number of layers increases with
increasing the width of the pore. The orientation of the
cyclohexane molecules is also studied. It is found that the
molecules at the outer layers lye almost parallel to the graphite
sheets, while in between of the layers the most preferable
orientation of the molecules is mainly perpendicular to the wall.
This behavior is similar to the one of the benzene molecules in
carbon nanotubes \cite{benzene-nanotube}.

In our previous publication \cite{benzen-me} we considered benzene
molecules in the same kind of graphite slit pores. It was shown
that at small $H$ the outer benzene layers crystallize in a planar
structure while the inner one stays relatively disordered. Unlike
this, all layers of cyclohexane behave qualitatively similar: all
layers simultaneously demonstrate freezing or melting transitions.
However, at large pore sizes $H$ the outer layers become more
ordered what is intuitively clear: the presence of a crystalline
wall with the same symmetry as the carbon ring of cyclohexane acts
as a periodic external potential on the liquid and makes it more
ordered.

From the local structure (Figs.~\ref{fig:fig4} and \ref{fig:fig5})
and behavior of the diffusion coefficient (Fig.~\ref{fig:fig6})
one can conclude that the system freezes with decreasing the width
of the pore and corresponding change of the number of layers from
$4$ to $3$. It is interesting that the freezing temperature of the
confined cyclohexane is higher than the bulk freezing temperature
in qualitative agreement with the experiment
\cite{mica01,mica2,mica02} on cyclohexane confined in the mica
pores. It should be noted that no signs of the hexatic phase were
found in the present study, however, further investigations in
this direction are necessary.

\subsection*{\sffamily \large ACKNOWLEDGMENTS}

%((Place Acknowledgments here))

The authors are grateful to S.M. Stishov and V.V. Brazhkin for
valuable discussions. Simulations were made at supercomputing
resources in NRC "Kurchatov Institute", which are supported as the
center for collective usage (project RFMEFI62114X0006, funded by
Ministry of Science and Education of Russia). The work was
supported by the Russian Science Foundation (Grant No
14-12-00820).

%((Additional Supporting Information may be found in the online version of this article.))

\clearpage

%%%%%%%%%%%%%%%%%%%%%%%%%%%%%%%%%%%%%%%%%%%%%%%%%%%%%%%%%%%%%%%%%%%%%%%%%%%%%%%%%
% BIBLIOGRAPHY

%\bibliography{bibtexrefs}   % Produces the bibliography via BibTeX.

%%%%%%%%%%%%%%%%%%%%%%%%%%%%%%%%%%%%%%%%%%%%%%%%%%%%%%%%%%%%%%%%%%%%%%%%%%%%%%%%%

\clearpage
%%%%%%%%%%%%%%%%%%%%%%%%%%%%%%%%%%%%%%%%%%%%%%%%%%%%%%%%%%%%%%%%%%%%%%%%%%%%%%%%%

%%%%%%%%%%%%%%%%%%%%%%%%%%%%

%%%%%%%%%%%%%%%%%%%%%%%%%%%%%%%%%%%%%%%%%%%%%%%%%%%%%%%%%%%%%%%%%%%%%%%%%%%%%%%%%
% FIGURE FILES

%\clearpage

%\vspace*{0.1in}   %%% FIGURE 1
%\begin{center}
%\includegraphics[width=0.2\columnwidth,keepaspectratio=true]{126.eps}
%\end{center}
%\vspace{0.25in}
%\hspace*{3in}
%{\Large
%\begin{minipage}[t]{3in}
%\baselineskip = .5\baselineskip
%Figure 1 \\
%Author A, Author B, Author C, Author D \\
%J.\ Comput.\ Chem.
%\end{minipage}
%}

%\clearpage

%\begin{table}
%\begin{tabular}{|c|c|c|c|}\hline
%\textbf{Quantity} & \textbf{Calculated} & \textbf{Observed} & \textbf{Error} \\ \hline
%  Density & 5.3 & 6.3 & Within limits \\ \hline
%  Optical magnification & 8.3 & 90.9 & Utterly unacceptable\! \\ \hline
%\end{tabular}
%\caption{\label{tbl1} Place table caption here.}
%\end{table}

\end{document}